\documentclass[aps,pra, twocolumn, superscriptaddress,nofootinbib]{revtex4-2}

\usepackage[dvipsnames]{xcolor}
\usepackage{hyperref}
\hypersetup{
  breaklinks=true,
  colorlinks=true,
  allcolors=BlueViolet,
}

\usepackage{graphicx} 
\usepackage{verbatim} 

\usepackage{physics,braket,bm}

\newcommand \be{\begin{equation}}
\newcommand \ee{\end{equation}}

\newcommand \bea{\begin{eqnarray}}
\newcommand \eea{\end{eqnarray}}

\newcommand{\diag}{\text{diag}}

\usepackage{amssymb}

\usepackage{booktabs}

\usepackage[inline]{enumitem}
\setlist[enumerate]{leftmargin=*}

\usepackage[normalem]{ulem}

\newcommand{\Strasbourg}
{University of Strasbourg and CNRS, CESQ and ISIS (UMR 7006), 67000 Strasbourg, France}
\newcommand{\InfleqtionM}{Infleqtion, Madison, WI 53703, USA}
\newcommand{\UWM}{Department of Physics, University of Wisconsin-Madison, 1150 University Avenue, Madison, WI 53706, USA}
\newcommand{\QPerfect}{QPerfect SAS, 23 rue du Loess, Strasbourg, France}

\newcommand{\IUF}{Institut Universitaire de France (IUF), 75000 Paris, France}

\definecolor{mscolor}{rgb}{0,0.5,0.5}
\definecolor{cpcolor}{rgb}{0.4,0,0.8}

\definecolor{sancolor}{rgb}{0.5,0,0.5}

\newcommand{\revision}[1]{\revision{#1}}

\newcommand {\rsub}[1]{\textcolor{black}{#1}}

\begin{document}

\title{Fault-tolerant quantum computation with static atomic buses}

\author{M. Bergonzoni}
\affiliation{\Strasbourg}

\author{L. Pecorari}
\affiliation{\Strasbourg}

\author{S. A. Norrell}
\affiliation{\UWM}

\author{C. Poole}
\affiliation{\UWM}

\author{G. Pupillo}
\affiliation{\Strasbourg}
\affiliation{\QPerfect}
\affiliation{\IUF}

\author{M. Saffman}
\affiliation{\UWM}
\affiliation{\InfleqtionM}
\date{\today}

\begin{abstract}

Efficient quantum error correction and fault-tolerant quantum computing require scalable, high-fidelity long-range connectivity. In neutral-atom quantum computers, this is commonly achieved through atom transport, but shuttling introduces latency and motional heating that worsen with system size. Here, we introduce a neutral-atom architecture 
based on static atomic buses, in which auxiliary mediator atoms enable long-range entangling operations without qubit transport. The architecture naturally supports long-range stabilizer measurements in high-rate LDPC codes and transversal logical gates between neighboring surface-code patches, enabling a modular framework for efficient logical memories, Clifford computation, and magic-state distillation. To realize these capabilities, we co-design optimal-control protocols for bus-mediated controlled-$Z$ gates that incorporate both microscopic neutral-atom dynamics and architectural constraints. 
We obtain smooth bus-mediated gates with fidelities approaching 99.9\% and durations of a few hundred nanoseconds by combining time-optimal control with interaction-flatness and robustness constraints. Large-scale simulations of quantum error correction and logical entangling operations between neighboring surface-code patches predict more than an order-of-magnitude improvement in logical error rates compared with atom-shuttling architectures under realistic noise. The architecture achieves logical gate times of approximately 100 $\mu$s and quantum-error-correction cycle times of about $1$ ms for code distances $d\lesssim12$. These results establish static atomic buses as a practical alternative to atom shuttling for scalable fault-tolerant neutral-atom quantum computing. 

\end{abstract}

\maketitle

\section{Introduction}

Quantum error correction (QEC) protects quantum information by delocalizing it across several entangled qubits, provided physical error rates remain below a threshold value \cite{Calderbank_1996}. Achieving universal fault-tolerant quantum computation (FTQC) requires preserving logical states within quantum memories and executing logical operations inside and across code blocks.
 
To fulfill the demanding requirements of FTQC, neutral-atom arrays have emerged as a leading platform because of long coherence times, flexible qubit connectivity, scalability, and high-fidelity operations. Fast controlled-$Z$ (CZ) gates based on the Rydberg-blockade mechanism have been demonstrated between adjacent atoms separated by $2.5$--$6.0\,\mu$m, with infidelities below $10^{-3}$ \cite{Evered2023,Radnaev2025,Tsai2025,Peper2025,Muniz2025a,Evered2026,Liu2026}, well below typical quantum-error-correction thresholds that are generally of the order of $10^{-2}$. Scaling these operations to the logical level, however, introduces severe connectivity challenges.
Specifically, high-efficiency quantum memories based on high-rate quantum low-density parity-check (LDPC) codes necessitate non-local operations for stabilizer readout. 
Executing entangling operations between code blocks
relies on logical transversal gates, which operate in $\mathcal{O}(1)$ time \cite{Zhou_2025,cain2025fastcorrelateddecodingtransversal,sk5y-25b1} but require a degree of connectivity that scales linearly with the linear size of the system,  $\mathcal{O}(d)$, with code distance $d$. 

The alternative to transversal operations is lattice surgery, which requires only nearest-neighbor connectivity but demands $\mathcal{O}(d)$ rounds of stabilizer measurements. However, dealing with many measurement rounds is often inconvenient in atomic platforms, where non-destructive mid-circuit readout is among the slowest operations, and so transversal operations are typically preferred. 
 
A leading approach to long-range connectivity is atom transport \cite{Bluvstein2022}, consisting of shuttling distant qubits into gate range, executing entangling operations locally, and returning them to their original positions. While this strategy has enabled pioneering experimental demonstrations of logical computation~\cite{Bluvstein2024,Bluvstein2026}, the time required to move atoms typically scales as $\mathcal{O}(d^{1/3})$~\cite{Rines2025}, adding latency that grows with code size. More critically, transport induces motional heating: atoms acquire vibrational excitations during acceleration, leading to dephasing errors whose infidelity scales as $\mathcal{O}(d^{2/3})$. 

A natural alternative to transport is to exploit the Rydberg interaction directly between distant static qubits. The van der Waals interaction between atoms spatially separated by $R$ scales as $\propto R^{-6}$, supporting high-fidelity blockade-based CZ gates up up to distances of a few $\mu$m \cite{Jandura2022}. 
This range can, in principle, be extended by employing larger principal quantum numbers $n$~\cite{Poole2025a,Pecorari2025}, applying DC electric fields or microwave dressing to reduce the Förster defect of dipole-coupled channels~\cite{Ashkarin_2025,Kurdak_2025}, or exploiting long-range resonant dipole--dipole interactions~\cite{Bergonzoni2026}. These approaches typically yield only moderate extensions of the interaction range, up to a few tens of $\mu$m, while significantly increasing the sensitivity to fluctuations in the resonance-inducing fields, background electric fields, and atomic positions.

A potentially scalable extension consists of using auxiliary atoms as quantum mediators, forming an \emph{atomic bus}. To build intuition on the advantages of this approach for van der Waals interactions, we observe that a chain of $N$ mediators divides the inter-atomic separation $R$ into $N+1$ segments of length $R/(N+1)$, potentially resulting in an effective nearest-neighbor interaction enhancement of $(N+1)^6$. This paradigm has been studied across many qubit platforms~\cite{Friesen2007} and explored theoretically for Rydberg systems~\cite{Wuster2010, Cesa2017, YSun2024, Doultsinos2025, Delakouras2026, He2024}, with an experimental demonstration in the context of maximum independent set problems~\cite{MKim2022}. Despite this progress, whether atomic buses can satisfy the demanding requirements of fault-tolerant quantum computing has remained an open question. Beyond enabling long-range interactions, a practical FTQC architecture requires gate protocols that achieve below-threshold fidelities in the presence of typical experimental noise, and integrate seamlessly with the architectural constraints of quantum error correction. 
These requirements have yet to be addressed within a unified architecture for fault-tolerant quantum computing.
 
In this work, we present an architecture that leverages static atomic buses to mediate long-range entanglement and natively supports all the necessary digital building blocks of a universal fault-tolerant quantum computer. First, it supports high-rate quantum LDPC memories by employing a stabilizer readout scheme, where bus ancillas mediate long-range stabilizer gates over tens of micrometer distances. Second, it naturally accommodates transversal entangling operations between adjacent code patches, such as surface codes. This enables a modular architectural design interleaving high-rate logical memories with lower-rate surface code bridges, which serve to efficiently compute the full logical Clifford gate set transversally, \rsub{requiring a number of operations scaling as $\mathcal{O}(d)$, each of $\mathcal{O}(1)$ duration, albeit at the cost of reduced parallelism}.
The surface code gate set can then be augmented to universality via magic state injection and distillation, which ultimately rely on non-Clifford physical operations and nonlocal transversal operations between code patches. The static atomic buses approach significantly suppresses motion-induced heating and operational latency. Active atom transport is here strictly limited to continuous qubit reloading and the routing of states from magic state factories. Furthermore, crosstalk-free, in-place mid-circuit measurements are achieved by employing a dual-species qubit encoding \cite{Anand2024,Miles2026}.

The first ingredient of our architecture is the rigorous design of optimal long-range entangling gates at the physical level. In particular, we co-design the gate pulses to explicitly incorporate both the microscopic interaction dynamics of the underlying neutral-atom hardware and the structural constraints imposed by quantum error correction at the logical level. In particular, the geometry of the atomic bus (i.e., the number of ancilla atoms and their spatial arrangement) is tailored to the quantum error-correcting code and the implemented operation (e.g., transversal logical gates or stabilizer measurements), while the physical parameters, such as the lattice spacing and the principal quantum numbers of the targeted Rydberg states, are optimized to maximize the gate fidelity for the given bus geometry.
Specifically, we use optimal-control techniques to find time-optimal pulses implementing high-fidelity CZ gates between distant qubits with the mediators fully disentangled at the end of the protocol. 
To mitigate motion-induced errors within a realistic dual-species architecture featuring rubidium and cesium atoms, we carefully select the Rydberg states and lattice spacing to create a locally flat interaction potential at the target atomic separation. This renders the gate intrinsically insensitive to small atomic displacements, suppressing motional infidelity down to the spontaneous-emission limit. In cases where this physical flatness cannot be achieved simultaneously for all relevant atom pairs, we explicitly incorporate robustness constraints against interaction variations directly into the pulse optimization. These pulse-engineering strategies enable bus-mediated gates with smooth pulse shapes, fidelities as high as $F\simeq0.999$, and durations of only a few hundred nanoseconds, comparable to state-of-the-art time-optimal nearest-neighbor gates~\cite{Jandura2022}. We design CZ gates suitable for transversal operations between surface-code blocks up to distance $d=11$ for parameters currently accessible in state-of-the-art experiments. Moreover, we show that the protocol can be extended to $d=25$ by employing higher Rydberg states and tighter lattice spacings. Although experimentally demanding, such parameters are within reach in light of recent proposals for suppressing crosstalk in tightly spaced arrays~\cite{Warttmann2026} and demonstrations of excitation to very high Rydberg states~\cite{Mohapatra2007,Sassmannshausen2013,Wang2017,Silpa2022}. \rsub{In this work,} similar gate durations and fidelities are obtained for non-local stabilizer operations in quantum LDPC codes.

At the logical level, to validate the viability of our gate protocols, we perform large-scale numerical simulations for long-range stabilizer readout with high-rate quantum LDPC codes (specifically, La-cross and bivariate bicycle codes), and transversal entangling gates between surface codes. These two processes constitute the foundational mechanisms driving high-efficiency quantum memory, fault-tolerant multi-qubit Clifford operations, and magic state distillation. Furthermore, we benchmark our static architecture for non-local syndrome extraction and transversal gates against an equivalent architecture that relies on dynamic atom shuttling. Our results demonstrate that, within noise parameter regimes currently accessible in state-of-the-art neutral atom devices, the static bus architecture yields over an order of magnitude improvement in performance compared to shuttling-based approaches, which are 
bottlenecked by motional heating effects. Within this architecture, we predict logical transversal gates in $\sim100\,\mu$s and QEC cycle times in the  $1$ ms regime for code distances $d\lesssim12$. All these results significantly improve upon the typical millisecond timescales of shuttling-based architectures.

These results establish static atomic buses as a practical, fast, and high-fidelity alternative to atom shuttling for satisfying the long-range connectivity requirements of fault-tolerant quantum computing. By explicitly restricting active atom shuttling to background operations, such as continuous qubit replenishment and the routing of resources from magic state factories, this architecture effectively eliminates transport-induced latencies from the primary algorithmic clock cycle. Furthermore, this work provides the first comprehensive, quantitative assessment of logical performance for a static neutral-atom architecture across optimal gate design, noise modeling, gate scheduling, and error-correction simulation for multiple code families.

The remainder of this work is organized as follows. Section~\ref{sec:css_and_bus} reviews the CSS code families and introduce the atom-bus Hamiltonians, gate fidelity measures, and optimal-control methodology. Section~\ref{sec.operations_surface_code} presents the bus-mediated transversal CZ gate for surface codes. Section~\ref{sec.long_range_stabilizer_readout} applies the framework to La-cross and bivariate bicycle codes. Section~\ref{sec.universality} outlines how one can interface high-rate quantum LDPC code memories with surface code units for fault-tolerant quantum computation, and an outlook is given in Sec.~\ref{sec.discussion}.

\begin{figure*}
    \centering
    \includegraphics[width=\linewidth]{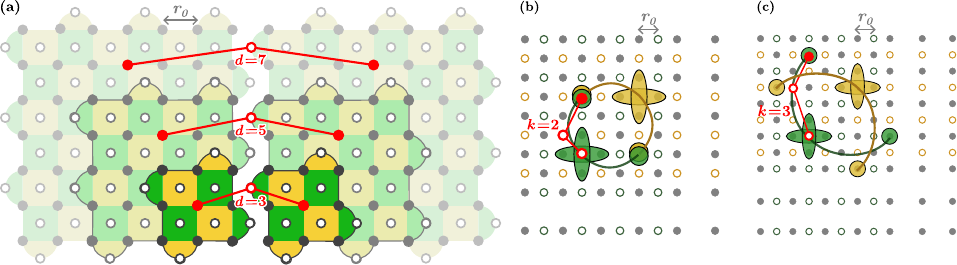}
    \caption{\textbf{(a)} Two surface-code patches of distance $d=3$, $5$, and $7$ (highlighted using different color gradings) with lattice spacing $r_0$. Black (white) dots represent data (ancilla) atoms. Red lines indicate examples of one-ancilla atomic buses enabling the implementation of CZ gates between corresponding data atoms in the two patches. \textbf{(b)} A patch of the $k=2$-La-cross code. The red line represents an example of a one-ancilla atomic bus enabling the implementation of a nonlocal stabilizer, i.e., a CZ gate between a data atom and an ancilla atom. \textbf{(c)} A patch of the $k=3$-La-cross code. The red line represents an example of a one-ancilla atomic bus enabling the implementation of a nonlocal stabilizer, i.e., a CZ gate between a data atom and an ancilla atom.}
    \label{fig:codes}
\end{figure*}

\section{CSS Codes and long-range atom-bus operations \label{sec:css_and_bus}}

In this work, we design a neutral atom architecture that leverages fast, long-range entanglement mediated by an atomic bus for quantum error correction. Because our proposed applications rely inherently on architectures using Calderbank–Shor–Steane (CSS) codes, in Sec.~\ref{sec.css_codes}, we begin by formally reviewing the CSS code framework. In Sec.~\ref{sec.gates},  we introduce the theoretical tools---including the general Hamiltonians, definitions for atom-bus mediated CZ gates and fidelities, and gate optimization tools---that are used to derive optimal pulses for error correction and computing in the
subsequent sections. Utilizing this unified framework, in the following sections we then demonstrate the utility of the atomic bus approach through two primary use cases: executing transversal entangling operations between isolated surface codes, and performing long-range syndrome extraction for high-rate quantum LDPC codes. 

\subsection{CSS Codes \label{sec.css_codes}}

CSS codes are stabilizer codes with stabilizer generators which are the tensor product of either only $X$ or only $Z$ Pauli operators \cite{Calderbank1996,steane-1996}. Their parity-check matrix $H\in\mathbb{F}_2^{M\times 2N}$ has the simple block anti-diagonal (symplectic) form
\begin{equation*}
    H_Q=\begin{pmatrix}
        0 & H_Z\\
        H_X & 0
    \end{pmatrix},
\end{equation*}
where $H_X,H_Z\in\mathbb{F}_2^{M\times N}$ are the binary matrices describing the action of the $M$ (not necessarily all linearly independent) $X$ and $Z$ stabilizers on the $N$ physical qubits, respectively. Since stabilizers commute, it must hold $H_XH_Z^T=0$. CSS codes are constructed from two classical linear codes $\mathcal{C}_1=[n_1,k_1,d_1]$ and $\mathcal{C}_2=[n_2,k_2,d_2]$ that must satisfy the nesting condition $\mathcal{C}_2\subseteq\mathcal{C}_1$ to ensure the stabilizer commutativity. 

CSS codes are particularly desirable for fault-tolerant quantum computing architectures because they correct separately for bit- and phase-flip errors, which significantly simplify the design of syndrome extraction circuits, and naturally support large sets of logical gates that can be implemented transversally. In the following, we review three quantum error correcting code families, namely surface, La-cross and bivariate bicycle codes, along with their connectivity requirements.

\subsubsection{Surface codes} 

In this work, we consider the problem of entangling two adjacent surface codes using static atom-bus interactions. For this purpose, we consider distance-$d$ \emph{rotated} surface codes \cite{PhysRevA.76.012305} that each use $d^2$ data qubits and $d^2-1$ ancilla qubits to encode a single logical qubit \cite{Fowler2012}. 
Logical operators are nontrivial strings of Pauli operators connecting opposite boundaries of the lattice, while error correction is performed through repeated syndrome extraction using ancillary qubits coupled locally to the data qubits with suitable gate ordering.
Owing to its strictly local connectivity, high threshold under circuit-level noise, and compatibility with nearest-neighbor architectures, the surface code has become the standard reference code for quantum error correction and fault-tolerant quantum computing.

The surface code \cite{Kitaev_2003,Bravyi1998,Fowler2012} is a CSS code constructed as the hypergraph product of two copies of a classical repetition code $\mathcal{C}_1=\mathcal{C}_2=[n,1,n]$ such that $H_X=(\mathbb{I}_{n\times n}\otimes H\quad H^T\otimes\mathbb{I}_{(n-1)\times(n-1)})$ and $H_Z=(H\otimes\mathbb{I}_{n\times n}\quad\mathbb{I}_{(n-1)\times(n-1)}\otimes H^T)$, with $H$ classical parity-check matrix of such repetition code. The surface code parameters, in its rotated configuration, are then $[[d^2,1,d]]$.

The surface code has local connectivity, with stabilizers acting only on four neighboring data qubits. Nevertheless, entangling logical operations, such as a transversal CZ gate, between surface codes requires physical CZ
gates pairwise coupling the corresponding data qubits in different code blocks. This corresponds to a long-range connectivity that scales with the system size, $d$. An example layout of surface code is shown in Fig.~\hyperref[fig:codes]{\ref{fig:codes}(a)}.

\subsubsection{La-cross codes} 

As first example of QEC code family with higher encoding rate than the surface code, we consider La-cross quantum low-density parity-check (LDPC) codes.

La-cross codes \cite{Pecorari2025} are also CSS hypergraph product \cite{tillich-2014,Kovalev_2013} codes constructed from two copies of the same cyclic code $\mathcal{C}=[n,k,d]$. To achieve higher encoding rates than the surface code, we require $k>1$, which is achieved by requiring $\mathcal{C}$ to have weight-$3$ checks to generalize the repetition code, which instead has weight-$2$ checks.  
Let $H\in\mathbb{F}_2^{(n-k)\times n}$ be the parity-check matrix of such classical code, for La-cross code one has $H_X=(H\otimes\mathbb{I}_{n\times n} \quad \mathbb{I}_{(n-k)\times(n-k)}\otimes H^T)$ and $H_Z=(\mathbb{I}_{n\times n}\otimes H \quad H^T\otimes\mathbb{I}_{(n-k)\times(n-k)})$. This construction yields a well-defined La-cross code with code parameters $[[N=n^2+(n-k)^2, K=k^2,D=d]]$ and open boundary conditions \cite{Pecorari2025,tillich-2014}. The stabilizers are weight-$6$ operators as a result of the product of two classical codes with weight-$3$ checks. These six data atoms are entangled with a single ancilla qubit via six entangling gates without incurring hook errors \cite{Pecorari2025}. Four of these six entangling gates are nearest-neighbor, while two are long-range with gate extent parametrized by $k$, which can be implemented statically using atom-bus gates. Having open boundary conditions is highly advantageous for experimental implementation, as it circumvents the need for additional non-local connectivity at the boundaries which would scale with the system size. Instead, for La-cross codes the long-range connectivity is fixed and scales as $\mathcal{O}(k)$, independently of the code distance. Example layouts of La-cross code are shown in Figs.~\hyperref[fig:codes]{\ref{fig:codes}(b)} and \hyperref[fig:codes]{\ref{fig:codes}(c)}. 

\subsubsection{Bivariate bicycle codes}

Another well-known family of high-rate CSS quantum LDPC codes is the bivariate bicycle code family \cite{Bravyi2024}. Let $I_\ell,S_\ell\in\mathbb{F}_2^{\ell\times\ell}$ be the identity matrix and the cyclic shift matrix, respectively, for some integer $\ell$. For bivariate bicycle codes $H_X=(A\quad B)$ and $H_Z=(B^T \quad A^T)$ where $A=A_1+A_2+A_3$ and $B=B_1+B_2+B_3$ are matrix trinomials with $A_i$ and $B_i$ powers of $x=S_\ell\otimes I_m$ and $y=I_m\otimes S_\ell$, respectively. This construction yields a well-defined bivariate bicycle code with code parameters $[[N=2\ell m, K=2\times\text{dim}(\text{ker}(A)\cap\text{ker}(B)), D]]$ with $ D=\text{min}\{|v|,\;v\in\text{ker}([A|B])\,\backslash\,\text{rowspace}([B^T|A^T])$ which can be computed using the integer linear programming method \cite{Bravyi2024}. The stabilizers are weight-$6$ operators acting on six data atoms entangled with a single ancilla qubit via six entangling gates. Two of these gates are long-range and can be implemented statically using atom-bus gates. Bivariate bicycle codes---contrary to La-cross codes---have periodic boundary conditions and hence boundary connectivity that would scale with the linear size of the code. To circumvent this problem and restore a boundary connectivity that does not scale with the code distance, we consider a technique known as \emph{array folding}~\cite{Poole2025a}, which we discuss in greater detail in the gate scheduling section~\ref{sec.bb_scheduling} for the specific example of code instance $[[144,12,12]]$ (Gross code).
An example layout of bivariate bicycle code is shown in Fig.~\ref{fig:bb_fig}.

\subsection{Atomic bus operations \label{sec.gates}}

The CSS codes described above rely on the ability to realize long-range gates in order to implement inter-code transversal logical operations and intra-code syndrome extraction. In the following sections, we introduce the theoretical tools---Hamiltonians, atomic-physics considerations, and optimal-control techniques---used to design such long-range gates through an atomic-bus approach, without the need to shuttle atoms.

\subsubsection{Atomic bus Hamiltonian \label{sec.Hamiltonian}}

Our goal is to realize entangling gates between two atoms located at the end points of a chain of length $(n_A+2)$, where $n_A$ ancilla atoms act as mediators in an atomic ``bus" configuration \cite{Friesen2007,Wuster2010,Cesa2017,YSun2024,Doultsinos2025,Pecorari2025b,Delakouras2026}. For later use in realizing transversal CNOT gates between two different surface codes and non-local stabilizers in qLDPC codes, respectively, we consider  two bus configurations: (i) two data atoms at the bus ends (at positions $i=0$ and $n_A+1$) [see Fig.~\hyperref[fig:codes]{\ref{fig:codes}(a)}]; or (ii) one data atom at one end ($i=0$) and an ancilla atom at the other end ($i=n_A+1$) [see Fig.~\hyperref[fig:codes]{\ref{fig:codes}(b,c)}]. Examples of configurations (i) and (ii) are discussed in Secs.~\ref{sec.operations_surface_code} and \ref{sec.long_range_stabilizer_readout}, respectively. Data and ancilla atoms are chosen to be different atomic species, specifically rubidium and cesium, respectively. Such a dual-species architecture is particularly attractive for static neutral-atom quantum computing, as it strongly suppresses crosstalk during gate operations, state preparation, measurements, and qubit reset, eliminating the need for atom shuttling during these operations.

Each data atom $D$ is modeled as a three-level system $\{\ket{0_D},\ket{1_D},\ket{r_D}\}$, where the qubit states $\{\ket{0_D},\ket{1_D}\}$ are encoded in hyperfine ground states and $\ket{r_D}$ is a Rydberg state used to mediate interactions [see Fig.~\hyperref[fig:1ancilla_gate]{\ref{fig:1ancilla_gate}(a)}]. Ancilla atoms $A$ are described either by the two-level structure $\{\ket{1_A},\ket{r_A}\}$ when acting as mediators, or by the three-level structure $\{\ket{0_A},\ket{1_A},\ket{r_A}\}$ when located at the ends of the bus [configuration (ii)] [see Fig.~\hyperref[fig:1ancilla_gate]{\ref{fig:1ancilla_gate}(a)}]. Since data and ancilla atoms are different atomic species, their internal states  differ. The total Hilbert-space dimension is therefore $\dim\left(\mathcal{H}\right)=3^2\times2^{n_A}$.

The total Hamiltonian of the system reads
\begin{equation}
    \label{eq:H_tot}
    H = \sum_{i=0}^{n_A+1} H_{\alpha_i}^{(i)} + H_{\rm int},
\end{equation}
where $H_{\alpha_i}^{(i)}$ denotes the interaction of atom $i$, of species $\alpha_i$, with an external global laser field, while $H_{\rm int}$ describes the van der Waals (vdW) interactions between atoms. Here the label $\alpha_i$ denotes the species of the $i$-th atom: in configuration (i), $\alpha_i = D$ for $i = 0, n_A + 1$ and $\alpha_i = A$ otherwise; in configuration (ii), $\alpha_i = D$ for $i = 0$ and $\alpha_i = A$ otherwise.

Two independent global laser drives, $\Omega_{\alpha_i}(t) = |\Omega_{\alpha_i}(t)| e^{i\phi_{\alpha_i}(t)}$, with $\alpha_i = D, A$, act on the data and ancilla atoms, respectively, coupling the qubit state $\ket{1_{\alpha_i}}_i$ to the Rydberg state $\ket{r_{\alpha_i}}_i$. \rsub{By global drives we mean that all atoms of the corresponding species in the bus are driven with the same Rabi frequency amplitude, exciting them to the same Rydberg state. This can be done with focused beams targeting specific bus atoms using, for example, the optical control architecture presented in \cite{Graham2023}. } The Rabi frequencies $\Omega_{\alpha_i}(t)$ are  taken to be complex, encoding both the amplitude \(|\Omega_{\alpha_i}(t)|\) and the phase \(\phi_{\alpha_i}(t)\) of the laser---with the laser detuning \(\Delta_{\alpha_i}(t)\) and the laser phase  related through $\Delta_{\alpha_i}(t)= d\phi_{\alpha_i}(t)/dt$.
The corresponding single-qubit Hamiltonian for species $\alpha$ and atom $i$ is
\begin{equation}
    \label{eq:laser_drive}
    H_{\alpha_i}^{(i)}(t) = \frac{\Omega_{\alpha_i}(t)}{2} \ket{r_{\alpha_i}}\bra{1_{\alpha_i}}_i + \mathrm{H.c.}
\end{equation}
with $\hbar=1$.

When atoms are excited to the Rydberg manifold, they interact via vdW interactions $B \propto R^{-6}$, where $R$ is the interatomic distance [See App.~\ref{app.interactions}]. The interaction potential also depends on the atomic species and principal quantum numbers of the interacting Rydberg states. The interaction Hamiltonian reads
\begin{equation}
    \label{eq:interaction}
    H_{\rm int} = \sum_{0 \le j < i \le n_A+1} B_{\alpha_i\alpha_j}^{i,j}\ket{r_{\alpha_i} r_{\alpha_j}}\bra{r_{\alpha_i} r_{\alpha_j}}_{i,j}.
\end{equation}
We use the notation $\ket{r_{\alpha_i} r_{\alpha_j}}\bra{r_{\alpha_i} r_{\alpha_j}}_{i,j} \equiv \ket{r_{\alpha_i}}\bra{r_{\alpha_i}}_i \otimes \ket{r_{\alpha_j}}\bra{r_{\alpha_j}}_j$, and $B_{\alpha_i\alpha_j}^{i,j}$ denotes the interaction strength between atoms $i$ and $j$, of species $\alpha_i$ and $\alpha_j$, separated by a distance $R_{\alpha_i\alpha_j}^{i,j}$. 

Conventional short-range gates on the scale of a few-$\mu$m are generally based on the blockade mechanism \cite{Levine2019, Jandura2022, Pagano2022, Mohan2023}, in which the interaction potential is much larger than the Rabi frequencies, $B_{\alpha_i\alpha_j}^{i,j} \gg |\Omega_{\alpha_i}|, |\Omega_{\alpha_j}|$, thereby preventing the simultaneous Rydberg excitation of two nearby atoms. In this work, we aim to realize long-range gates in the range of tens of $\mu$m using an atomic bus. We find below that \rsub{increasing the number of mediators reduces the gate fidelity, primarily because it increases the total time spent in the Rydberg manifold and thus the exposure to spontaneous emission. The stronger mediator–mediator interactions do not compensate for this effect, as the gate duration remains finite and the additional next-nearest-neighbour interactions further complicate the dynamics. Therefore,} it is advantageous to reduce the number of ancilla atoms in the bus as much as possible. As a consequence, the atoms generally lie outside each other’s blockade radius, with interaction strengths typically of the order of $1$--$5$ times the Rabi frequency. This makes the gate sensitive to the precise values of the interaction strengths $B_{\alpha_i\alpha_j}^{i,j}$. However, sensitivity to precise values of the interaction can be mitigated by appropriately choosing the Rydberg states of the two atomic species [see Sec.~\ref{sec.gate_1ancilla} below] and using quantum optimization tools, as explained below.

We note that the full system dynamics determined by Eq.~\eqref{eq:H_tot} with Hilbert-space dimension $3^2 \times 2^{n_A}$ can be studied by focusing on only three invariant subspaces of dimensions $2^{n_A}$, $2^{n_A+1}$, and $2^{n_A+2}$, corresponding to the number of atoms at the ends of the bus prepared in the state $\ket{1_{\alpha_i}}$ being zero, one, or two, respectively. See App.~\ref{app.subspaces} for more details.

\subsubsection{Bus-mediated long-range CZ gates}

We aim at implementing entangling gates  between the two qubits at the ends of the chain by mediating the interaction through the $n_A$ ancilla atoms forming the bus, while ensuring that the latter are fully disentangled from the end qubits at the conclusion of the protocol. 

Given global laser pulses \(\Omega_{\alpha_i}(t)\) of duration \(T\), we denote by
$
U(T)=\tau\exp\left(-i\int_0^T H(t)\,dt\right)
$
the time-evolution operator, with $H(t)$ the Hamiltonian of Eq.~\eqref{eq:H_tot}. The pulse implements a phase gate between the two end qubits of the chain if
\begin{equation}
    \label{eq:phasegate}
U(T)\ket{q}=e^{i\xi_q}\ket{q},
\end{equation}
where $q$ belongs to the computational basis of the two end atoms, $q=(q_0,q_{n_A+1})\in Q=\{00,01,10,11\}$, and $\xi_q$ is the phase acquired by the state $\ket{q}$. A CZ phase gate corresponds to the condition \cite{Jandura2022}
\begin{equation}
    \xi_{00}-\xi_{01}-\xi_{10}+\xi_{11}=\pi.
\end{equation}
Including the $n_A$ ancilla bus-atoms, the gate reads
\begin{align}
    \nonumber
    \mathrm{CZ} \ket{q_0\,1\ldots1\,q_{n_A+1}} \to \ket{q_0\,1\ldots1\,q_{n_A+1}}\\
    \label{eq:CZ}
     \times\exp\left[i\pi q_0q_{n_A+1} + i\theta(q_0+q_{n_A+1})\right],
\end{align}
where $q_0,q_{n_A+1} \in \{0,1\}$ denote the states of the two end qubits and $\theta$ is the phase associated with a final global phase gate $\diag(1,e^{i\theta})$. For simplicity, the atomic-species labels $\alpha_i = D, A$ have been omitted. The ancilla mediators are required to begin and end the evolution in the state $\ket{1_A}$: since the ancilla atoms are introduced specifically to mediate the interaction, there is no advantage in initializing them in states decoupled from the Rydberg manifold.

To benchmark the performance of a gate $U(T)$, for a given choice of  $\Omega_{\alpha_i}(t)$ and pulse duration $T$, in reproducing the target unitary $\mathrm{CZ}$, we introduce the Bell-state fidelity $F$, defined as~\cite{Levine2019, Graham2019, Theis2016b, Robicheaux2021, Jandura2022}
\begin{equation}
    \label{eq:1-F}
    F= \frac{1}{16} \left| \sum_{q \in Q} \bra{q} \mathrm{CZ}^\dagger P(\theta)^\dagger U(T) \ket{q} \right|^2,
\end{equation}
with $P(\theta)$ a global phase gate [see Eq.~\eqref{eq:CZ}]. Equation \eqref{eq:1-F} measures the quality of a pulse through the fidelity between the state obtained by applying the pulse to
$\ket{++}=\left(\ket{00}+\ket{01}+\ket{10}+\ket{11}\right)/2$
and the target output state $\mathrm{CZ}\ket{++}$, up to the phase gate $P(\theta)$. In this work, we are generally interested in time-optimal pulses, defined as the shortest pulses that achieve maximal fidelity, $F = 1$, with $F$ from  Eq.~\eqref{eq:1-F}.  

We note that the fidelity expression in Eq.~\eqref{eq:CZ} accounts only for the coherent evolution induced by the laser pulses $\Omega_{\alpha_i}(t)$, while the effects of dominant error sources---namely spontaneous emission and thermal motion---are not included. In Sects. \ref{sec.operations_surface_code} and \ref{sec.long_range_stabilizer_readout} below, we provide both analytical estimates and quantitative measures of the impact of these noise sources on gate fidelities via numerical simulations for each given time-optimal pulse solution. Wherever necessary, we provide optimized solutions to counter the dominant noise sources. 

\subsubsection{Gate fidelity optimization}

In order to design laser pulses $\Omega_{\alpha_i}(t)$ that maximize the gate fidelity $F$, we use the quantum optimal-control method known as Gradient Ascent Pulse Engineering (GRAPE). GRAPE optimizes a set of control fields $u(t)$ entering the Hamiltonian $H$ by minimizing a cost functional $\mathcal{C}[u]$ through gradient-based optimization~\cite{Khaneja2005, Garon2013, Riaz2019, wilhelm2020, Jandura2022, Jandura2023, Bergonzoni2026}. The time-dependent controls are parametrized using a piecewise-constant ansatz, $u(t)=u_j$ for $t\in[j\Delta t,(j+1)\Delta t]$, with $\Delta t=T/N$ and $N$ the number of time intervals, reducing the problem to an optimization over the parameters $\{u_0,\ldots,u_{N-1}\}$. A key feature of GRAPE is the efficient evaluation of the gradients $\partial \mathcal{C}/\partial u_j$, enabling the use of optimization algorithms such as the Broyden--Fletcher--Goldfarb--Shanno (BFGS) method~\cite{scipy2020}.

\begin{figure*}
    \centering
    \includegraphics[width=0.81\linewidth]{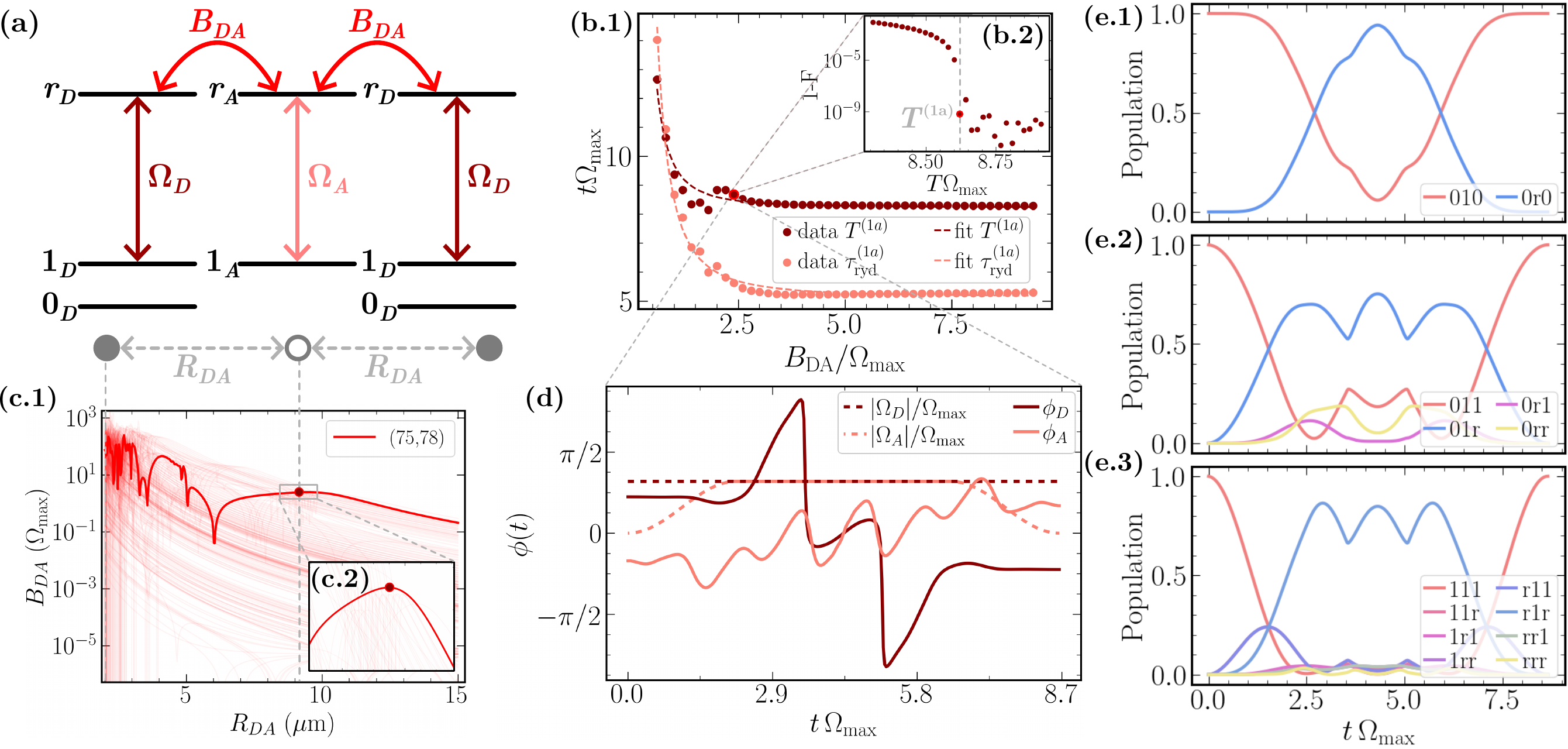}
    \caption{One-mediator bus gate for transversal operations between surface codes of distance $d=7$, assuming $\Omega_{\rm max}=2\pi\times 2\,\mathrm{MHz}$. \textbf{(a)} Level scheme of a three-atom bus in configuration (i), consisting of two data atoms $D$ and one ancilla atom $A$ acting as a mediator. Nearest neighbors interact with strength $B_{DA}$ and are driven by independent laser fields $\Omega_{\alpha_i}$, with $\alpha_i=D,A$. \textbf{(b.1)} Gate duration $T^{(1a)}$ and time spent in the Rydberg manifold $\tau_{\rm ryd}^{(1a)}$ for time-optimal pulses, shown as functions of $B_{DA}$. Numerical optimization results (dots) are fitted using Eqs.~\eqref{eq:T_gate_1a} and \eqref{eq:tau_ryd_1a} (dashed lines). \textbf{(b.2)} Infidelity $1-F$ of the GRAPE pulses as a function of the pulse duration $T$, for $B_{DA}/\Omega_{\rm max}=2.48$. The time-optimal pulse is the shortest one achieving numerical zero infidelity. \textbf{(c)} Data--ancilla interaction potential $B_{DA}(R_{DA})$ for principal quantum numbers $(n_{\rm Rb}=75,n_{\rm Cs}=78)$. Inset \textbf{(c.2)}: zoom-in of the local maximum corresponding to the data--ancilla distance (vertical dashed line) for a lattice spacing $r_0=2.68\,\mathrm{\mu m}$. \textbf{(d)} Phases $\phi_{\alpha_i}$ and amplitudes $|\Omega_{\alpha_i}|$ of the time-optimal gate with $B_{DA}/\Omega_{\rm max}=2.48$. \rsub{\textbf{(e)} Evolution of the state populations during the pulse of Fig.~\hyperref[fig:1ancilla_gate]{\ref{fig:1ancilla_gate}(d)}, starting from the computational-basis states $\ket{010}$ in (e.1), $\ket{011}$ in (e.2), and $\ket{111}$ in (e.3) [see the subspace decomposition in App.~\ref{app.subspaces}].}}
    \label{fig:1ancilla_gate}
\end{figure*}

In our implementation, the control parameters are the laser phases $\phi_{\alpha_i}(t)$ and the global phase $\theta$ [see Eq.~\eqref{eq:CZ}], while the laser amplitudes are fixed to their maximum value, assumed identical for the two drives, i.e., $|\Omega_D(t)| = |\Omega_A(t)| = \Omega_{\rm max}$. We verified that fixing the amplitudes does not compromise time optimality: even when $|\Omega_{\alpha_i}(t)|$ are included among the optimized controls, GRAPE converges to solutions with constant amplitudes at their maximum values. Accordingly, we adopt this configuration throughout the work, except in a few cases where $|\Omega_A(t)|$ follows a Tukey profile [Fig.~\hyperref[fig:1ancilla_gate]{\ref{fig:1ancilla_gate}(d)}]. While looking for time-optimal pulses, as cost functional $\mathcal{C}_{\rm to}$ we use the Bell-state infidelity derived from Eq.~\eqref{eq:1-F}, namely 
\begin{equation}
    \mathcal{C}_{\rm to}[\phi_{\alpha_i},\theta]=1-F.
\end{equation}
\rsub{Note that the optimization excludes spontaneous emission from the Hamiltonian, as time-optimal pulses generally also minimize the time spent in the Rydberg manifold. Its contribution to the infidelity is evaluated a posteriori using a non-Hermitian model of Rydberg decay.\\}
Different cost functionals can be used to include robustness against specific error sources in the optimization. For example, in atom-bus gates---when working far from the blockade regime---sensitivity to variations in the interaction strength $B$, e.g. due to atomic motion, is a major concern. This type of errors can be mitigated by carefully choosing the interacting pair of Rydberg states [see Sec.~\ref{sec.gate_1ancilla} below]. However, when this strategy is not feasible (for example in asymmetric buses), we adopt a sampling or ensemble approach \cite{Hughes2020, Goerz2011, Chen2014, Giudici2025}, modifying the cost functional to include robustness constraints directly in the optimization, with cost functional 
\begin{equation}
    \mathcal{C}_{\rm rob}[\phi_{\alpha_i},\theta] = \sum_{m=0}^M \big( 1 - F[B(1+\Delta B_m)] \big),
\end{equation}
where $F[B(1+\Delta B_m)]$ is the fidelity of the pulse for an interaction strength $B(1+\Delta B_m)$, and the $M$ relative variations $\Delta B_m$ are uniformly sampled in the interval $[-0.05,0.05]$.

\section{Transversal operations between surface codes\label{sec.operations_surface_code}}

We start the discussion of long-range operations in CSS codes by presenting a neutral-atom atomic-bus scheme for implementing transversal CZ gates between distance-$d$ surface codes using one or more ancilla atoms as mediators. In Sec.~\ref{sec.gate_1ancilla}, using the theoretical tools introduced above we derive  time-optimal pulses, investigate their performance in terms of speed and fidelity, and provide semi-analytical scalings of gate fidelities for relevant physical quantities such as the Rydberg principal quantum numbers $n$ and lattice spacing $r_0$. 
In Sec.~\ref{sec:bellpair} we discuss how to use these time-optimal gate protocols to implement a logical Bell pair between surface codes. In Sec.~\ref{sec:sc-noise} we present the hardware-aware noise model used in our numerical simulations and benchmark the performance of our static architecture against one using atom-rearrangement. 

\subsection{Long-range CZ gate between two data atoms \label{sec.gate_1ancilla}}

\subsubsection{Atomic bus with a single mediator }

Consider a three-atom bus consisting of two data atoms $D$ at the ends [configuration (i)] and a single ancilla atom $A$ located exactly in the middle [see Fig.~\hyperref[fig:1ancilla_gate]{\ref{fig:1ancilla_gate}(a)}]. The aim is to realize a transversal CZ gate between two surface codes of distance $d$. Given a lattice spacing $r_0$, the distance between data and ancilla atoms is $R_{DA} \equiv R_{DA}^{0,1} = R_{AD}^{1,2}  = r_0\sqrt{d^2+1}/2$ [Fig.~\hyperref[fig:codes]{\ref{fig:codes}(a)}].

From the geometry of this setup, the interaction strength is assumed to be finite and symmetric: the ancilla atom is equidistant from the two data atoms $B_{DA}\equiv B_{DA}^{0,1}=B_{AD}^{1,2} $ and lies outside their mutual blockade radius, so that $B_{DA}/\Omega_{\rm max} < \infty$. The full Hamiltonian in Eq.~\eqref{eq:H_tot} takes the form
\begin{equation}
    \label{eq:H_tot_1a}
    H(t) = H_{D}^{(0)}(t) + H_A^{(1)}(t) + H_{D}^{(2)}(t) + H_{\rm int}^{\rm (1a)},
\end{equation}
where the interaction Hamiltonian [see  Eq.~\eqref{eq:interaction}] reads
\begin{equation}
    \label{eq:interaction_1a}
    H_{\rm int}^{\rm (1a)} = B_{DA} \left(\ket{r_D r_A}\bra{r_D r_A}_{0,1} + \ket{r_A r_D}\bra{r_A r_D}_{1,2}\right),
\end{equation}
where the label $(\rm 1a)$ refers to a symmetric configuration with only one mediator: data-ancilla-data. For simplicity, the data-data interaction is assumed here negligible, i.e., $B_{DD}=0$; however, its presence is in any case not detrimental if it is directly included in the optimization.

In the following, atomic bus gates are optimized for finite values of the interaction strength $B_{DA}$ and are therefore sensitive to fluctuations induced by thermal motion of the atoms, with infidelity scaling approximately as
\begin{equation}
    \label{eq:1-F_mot}
    (1 - F)_{\rm mot} \approx \Delta B_{DA}^2 \tau_{\rm 2ryd}^2.
\end{equation}
Here, $\Delta B_{DA}$ denotes fluctuations of the blockade interaction and $\tau_{\rm 2ryd}$ is the average time during which two nearby atoms are simultaneously excited to Rydberg states [see App.~\ref{app.infidelity}]. These errors related to atomic motion can be strongly suppressed by choosing Rydberg-state pairs $(n_{\rm Rb}, n_{\rm Cs})$ and lattice spacings $r_0$ such that the interaction potential is locally flat near the equilibrium data--ancilla separation $R_{DA}$. Such flat regions arise naturally in heteronuclear systems because the two species possess different level structures and $R$-dependent coupling strengths, enabling cancellations that are typically suppressed for identical atoms in the same Rydberg state. An example is shown in Fig.~\hyperref[fig:1ancilla_gate]{\ref{fig:1ancilla_gate}(c.1)} for a given choice of principal quantum numbers $n_{\rm Rb}=75$ and $n_{\rm Cs}=78$, where the chosen interaction potential (thick red continuous line) exhibits a local maximum with 
\begin{equation}
    \label{eq:dB/dR=0}
    \left. \frac{d B_{DA}(R)}{d R} \right|_{R=R_{DA}} = 0,
\end{equation}
for $R_{DA} \simeq 9.5\, \mu$m [see Fig.~\hyperref[fig:1ancilla_gate]{\ref{fig:1ancilla_gate}(c.2)}]. \rsub{In such flat regions of the interaction potential, the dependence on the interatomic distance $R_{DA}$ is no longer given by the naive scaling $B_{DA}\propto R_{DA}^{-6}$; instead, the interaction is locally first-order insensitive to the distance. Thus, for small fluctuations of the interatomic distance $\Delta R_{DA}$, such as those induced by thermal motion, there is no appreciable variation of the potential, $\Delta B_{DA}=0$ [see Eq.~\eqref{eq:dB/dR=0}], and therefore no appreciable effect on the gate fidelity [see Eq.~\eqref{eq:1-F_mot}].}

In the following, for a fixed target separation $R_{DA}$, we scan pairs $(n_{\rm Rb}, n_{\rm Cs})$ in the range $40 \le n_{\rm Rb,Cs} \le 80$ and identify interaction potentials with a nearby local maximum at $R=R^*$ satisfying $B_{DA}(R^*) > \Omega_{\rm max}$. We then choose the lattice spacing $r_0$, ideally near $3\,\mu\mathrm{m}$ to simplify addressability in experiments, such that
\begin{equation}
    R_{DA} = \frac{r_0}{2}\sqrt{d^2+1} \approx R^* .
\end{equation}
This choice, whenever possible, renders the gates first-order insensitive to atomic motion.

\begin{figure*}
    \centering
    \includegraphics[width=1\linewidth]{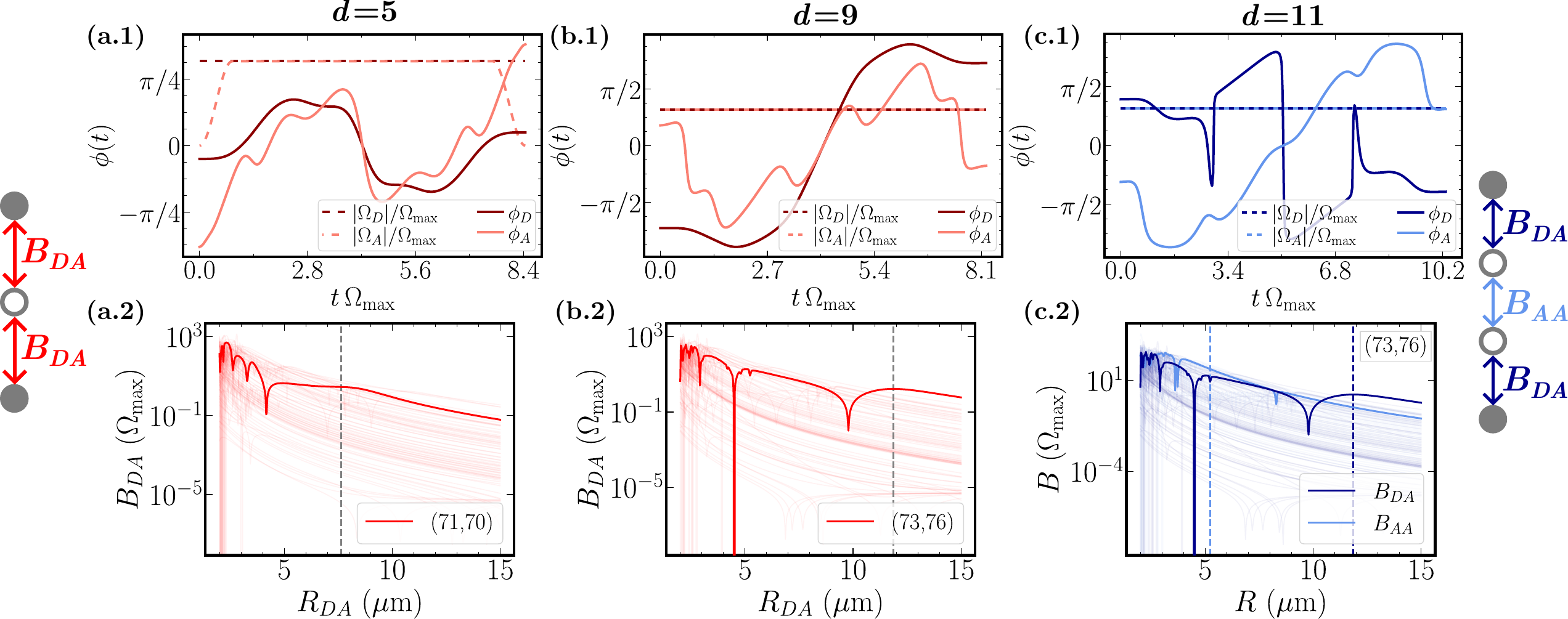}
    \caption{Transversal CZ gates between surface codes considering $\Omega_{\rm max}=2\pi\times 2\,\mathrm{MHz}$. \textbf{(a)} Phases $\phi_{D,A}$ and amplitudes $|\Omega_{D,A}|$ of the time-optimal gate for a $d=5$ surface code with $B_{DA}/\Omega_{\rm max}=2.85$ (a.1), together with the corresponding data--ancilla interaction potential for principal quantum numbers $n_{\rm Rb}=71$ and $n_{\rm Cs}=70$ and lattice spacing $r_0=3\,\mathrm{\mu m}$ (a.2). \textbf{(b)} Phases $\phi_{D,A}$ and amplitudes $|\Omega_{D,A}|$ of the time-optimal gate for a $d=9$ surface code with $B_{DA}/\Omega_{\rm max}=1.69$ (b.1), together with the corresponding data--ancilla interaction potential for principal quantum numbers $n_{\rm Rb}=73$ and $n_{\rm Cs}=76$ and lattice spacing $r_0=2.62\,\mathrm{\mu m}$ (b.2). \textbf{(c)} Phases $\phi_{D,A}$ and amplitudes $|\Omega_{D,A}|$ of the time-optimal gate for a $d=11$ surface code with $B_{DA}/\Omega_{\rm max}=1.69$ and $B_{AA}/\Omega_{\rm max}=41.95$ (c.1), together with the corresponding data--ancilla (dark line) and ancilla--ancilla (light line) interaction potentials for principal quantum numbers $n_{\rm Rb}=73$ and $n_{\rm Cs}=76$ and lattice spacing $r_0=2.62\,\mathrm{\mu m}$ (c.2).}
    \label{fig:gate_surface_code}
\end{figure*}

\subsubsection{Time-optimal bus-mediated CZ gates and relevant noise}

Time-optimal pulses implementing a CZ gate between the data atoms can be obtained for different finite values of the interaction strength $B_{\rm DA}/\Omega_{\rm max}$ using quantum optimization methods. This is exemplified in the inset of Fig.~\hyperref[fig:1ancilla_gate]{\ref{fig:1ancilla_gate}(b.2)} where, for a fixed value of $B_{DA}/\Omega_{\rm max} = 2.48$, GRAPE is shown to converge to pulses with numerically vanishing infidelity $1-F$ when the pulse duration $T$ exceeds a critical value, $T>T^{(1a)}$. Conversely, for $T<T^{(1a)}$, the algorithm is no longer able to find pulses that drive the cost functional to zero, and we observe a sharp jump in the infidelity. We therefore define the time-optimal pulse as the one obtained at the critical duration $T^{(1a)}$, i.e., the shortest pulse achieving numerically vanishing infidelity.

Figure~\hyperref[fig:1ancilla_gate]{\ref{fig:1ancilla_gate}(d)} shows the corresponding time-optimal pulses $\phi_{D,A}(t)$ and $|\Omega_{{D,A}}(t)|$ vs. time $t$, which implement the transversal CZ gate between surface codes of distance $d=7$, for a choice of realistic experimental  parameters---i.e. ($n_{\rm Rb}=75$, $n_{\rm Cs}=78$), Rabi frequency $\Omega_{\rm max}=2\pi\times 2\,\mathrm{MHz}$ and lattice spacing $r_0=2.68\,\mathrm{\mu m}$, chosen to reflect conservative values demonstrated in current experiments \cite{Evered2023,Radnaev2025,Tsai2025,Peper2025,Muniz2025a,Evered2026,Liu2026}. We find that the time-optimal pulses correspond to pulses where the $|\Omega_{D,A}|$ are kept essentially constant, while the phases are modulated in time in a continuous, smooth way. The  duration of the time-optimal pulse is found to be $T^{(1a)}\Omega_{\rm max}=8.65$, which is comparable to the state-of-the-art time-optimal pulses for nearest-neighbor interactions in the blockade limit $T^{{\rm (to)}}\Omega_{\rm max}=7.612$ \cite{Jandura2022}. \rsub{Figure~\hyperref[fig:1ancilla_gate]{\ref{fig:1ancilla_gate}(e)} shows the evolution of the population of each internal state along the time-optimal pulse in Fig.~\hyperref[fig:1ancilla_gate]{\ref{fig:1ancilla_gate}(d)}, starting from different computational-basis states.} The resulting computed infidelity is $1-F=2.8\times10^{-3}$ and is found to be dominated by spontaneous emission at room temperature, from both data and ancilla atoms.
We find that these pulse characteristics---i.e., phase continuity and smoothness, pulse duration, sensitivity to spontaneous emission---are common to all situations relevant for the investigated code distances $d \leq 9$. As examples, panels~\hyperref[fig:gate_surface_code]{(a)} and~\hyperref[fig:gate_surface_code]{(b)} of Fig.~\hyperref[fig:gate_surface_code]{\ref{fig:gate_surface_code}} present our results for time-optimal pulses for the cases $d=5$ and $d=9$, respectively, showing similar behavior. \rsub{A complete list of the pulses optimized for transversal CZ gates between surface codes, together with their parameters, is provided in Table~\ref{tab:all_surface_code_gates}.} We note that operating in a cryogenic environment can suppress blackbody-radiation-induced transitions from the Rydberg states \cite{CantatMoltrecht2020, Schymik2021, Pichard2024, ZZhang2025, JJin2026}, possibly further reducing the above gate infidelity. However, we note that throughout this work the reported infidelities do not include errors arising from laser phase and frequency noise, nor from stray electric fields that can perturb high-$n$ Rydberg states.

We further characterize the general behavior of time-optimal pulses in Fig.~\hyperref[fig:1ancilla_gate]{\ref{fig:1ancilla_gate}(b.1)}, where we plot the gate duration $T^{\rm(1a)}$ together with the time $\tau_{\rm ryd}^{\rm(1a)}$ spent in the Rydberg manifold for different values of the interaction strength in the range $B_{\rm DA}/\Omega_{\rm max} \in [0.1,10]$. The time spent in the Rydberg manifold, $\tau_{\rm ryd}$, is particularly important as it provides an estimate of the error budget due to spontaneous emission $ (1 - F)_{\rm ryd} \approx \Gamma \tau_{\rm ryd}^{\rm(1a)}$ where $\Gamma\sim n^{-3}$ is the decay rate of the chosen Rydberg states at room temperature [see App.~\ref{app.infidelity}]. 

We find that the data in Fig.~\hyperref[fig:1ancilla_gate]{\ref{fig:1ancilla_gate}(b.1)} are well fitted [dashed lines in Fig.~\hyperref[fig:1ancilla_gate]{\ref{fig:1ancilla_gate}(b.1)}] by the following expressions
\begin{equation}
    \label{eq:T_gate_1a}
    T^{\rm(1a)} \Omega_{\rm max} \sim 8.2 + \frac{1.5}{(B_{\rm DA}/\Omega_{\rm max})^{2}},
\end{equation}
\begin{equation}
    \label{eq:tau_ryd_1a}
    \tau_{\rm ryd}^{\rm(1a)} \Omega_{\rm max} \sim 5.2 + \frac{3.4}{(B_{\rm DA}/\Omega_{\rm max})^{2}}.
\end{equation} 
The explicit dependence of $\tau_{\rm ryd}^{\rm(1a)}$ on $B_{DA}/\Omega_{\rm max}$ in Eq. \eqref{eq:tau_ryd_1a} allows us to provide an analytic estimate of the CZ gate fidelity for generic choices of experimental parameters as follows: the van der Waals interaction $B_{DA}$ scales as $B_{DA}\sim n^{11}/(dr_0)^6$, with $n$ the principal quantum number, $d$ the surface-code distance, and $r_0$ the lattice spacing, while the Rabi frequency scales as $\Omega_{\rm max}\sim En^{-3/2}$, with $E$ the laser electric field, resulting in the ratio
\begin{equation}
    \frac{B_{DA}}{\Omega_{\rm max}}
    \sim
    \frac{n^{25/2}}{E(dr_0)^6}.
\end{equation}
Inserting this expression in Eq.~\eqref{eq:tau_ryd_1a}, we obtain the following gate infidelity due to spontaneous emission 
\begin{equation}
    \label{eq:1-F_scaling}
    1-F^{\rm(1a)} \sim \frac{1}{En^{3/2}} \left( c_0 + c_1\frac{E^2d^{12}r_0^{12}}{n^{25}}
    \right),
\end{equation}
where $c_0$ and $c_1$ are numerical constants [see App.~\ref{app:scalability} for more details]. The first term in the parenthesis corresponds to the ideal blockade regime, $B_{DA}/\Omega_{\rm max}\gg1$, while the second one describes the additional error due to longer-range gates with finite interactions, $B_{DA}\sim\Omega_{\rm max}$, which grows rapidly with $d$ and $r_0$.

Equation~\eqref{eq:1-F_scaling} indicates how the infidelity of the transversal CZ gate can be reduced. In particular, improvements can be achieved by accessing higher Rydberg states, with larger principal quantum numbers $n$, which enhance both the interaction strength and the Rydberg lifetime, and by increasing the laser power $E$, thereby allowing larger Rabi frequencies $\Omega_{\rm max}$. This assumes that laser intensity and phase stability are maintained and that external-field fluctuations do not significantly perturb the Rydberg states.
 As an example, we optimized a gate for a surface code of distance $d=7$ using $n_{\rm Rb}=n_{\rm Cs}=89$, $\Omega_{\rm max}=2\pi\times8\,\mathrm{MHz}$, and $r_0=2.65\,\mathrm{\mu m}$. The resulting pulse has a duration $T^{(1a)}\Omega_{\rm max}=8.44$ and an infidelity of $1-F=5.3\times10^{-4}$, that is largely below the code thresholds of the CSS codes considered in this work. This corresponds to an improvement by approximately a factor of five compared with the pulse shown previously [see Fig.~\hyperref[fig:1ancilla_gate]{\ref{fig:1ancilla_gate}(d)}].  Similar performances can be achieved also for the cases $d=5$ and $d=9$.

Equation~\eqref{eq:1-F_scaling} also clarifies how the gate scales with code distance $d$. Increasing the distance as $d\to\eta d$, with $\eta>1$, while maintaining approximately constant infidelity requires either reducing the lattice spacing as $r_0\to r_0/\eta$ or increasing the principal quantum number as $n\to n\,\eta^{24/53}$. Increasing the laser power $E$ instead produces a nonmonotonic effect: although it decreases the overall prefactor, it also reduces the ratio $B_{DA}/\Omega_{\rm max}$, thereby enhancing the second term in Eq.~\eqref{eq:1-F_scaling}.
As an illustrative example, the longest-range gate discussed above corresponds to $d=9$, $n\approx75$, and $r_0\approx3\,\mathrm{\mu m}$. According to the scaling analysis, extending the range to $d=25$ keeping the same infidelity $1-F$ would require approximately $n\approx95$ and $r_0\approx1.7\,\mathrm{\mu m}$. This prediction has been validated by explicitly optimizing a gate for a $d=25$ surface code using $n_{\rm Rb}=99$, $n_{\rm Cs}=96$, and $r_0=1.7\,\mathrm{\mu m}$. Nearest-neighbor gates with this lattice spacing can still be achieved while avoiding the spaghetti region by using lower values of $n$ \cite{Evered2026}. The resulting pulse has a duration $T^{(1a)}\Omega_{\rm max}=8.85$, an impressive coherent interaction range $>40\,\mu$m, and an infidelity of $1-F=3.6\times10^{-3}$. In this regime, the infidelity receives a significant contribution from atomic motion, since no sufficiently flat region of the interaction potential is available at the required interatomic distance. While experimentally challenging, such parameters are within future experimental reach, given existing proposals to suppress crosstalk between closely spaced atoms~\cite{Warttmann2026} and demonstrations of excitation to very high Rydberg states~\cite{Mohapatra2007, Sassmannshausen2013, Wang2017, Silpa2022}.

\rsub{We remark here that the scaling argument in Eq. \eqref{eq:1-F_scaling} is used only to provide a qualitative description of the scaling behavior of the infidelity as a function of the principal physical parameters. However, whenever this scaling argument is used to infer strategies for increasing the fidelity or extending the range (e.g., in the two paragraphs above), its validity is explicitly tested and confirmed by directly optimizing a pulse for the corresponding set of parameters.}\\

Alternatively, realizing entangling gates between two data atoms at distances $R > 25\,\mathrm{\mu m}$ corresponding to code distances $d \geq 11$, keeping $n_{\rm Rb, Cs}\leq80$ and $r_0\approx3\,\mathrm{\mu m}$, requires adding one or more additional ancilla atoms to the bus to bridge the interaction [see Fig.~\hyperref[fig:gate_surface_code]{\ref{fig:gate_surface_code}(c)}]. The introduction of additional atoms complicates the system dynamics by introducing additional energy scales, namely the ancilla–ancilla interaction strength $B_{AA}$, and the next-nearest neighbor interactions, e.g., $B_{DA}^{0,i}$ with $i\ge2$. The extra atoms also increase the global time spent in the Rydberg manifold, thereby enhancing the effects of spontaneous emission. Moreover, the presence of two compatible flat regions in the interaction potentials $B_{DA}(R)$ and $B_{AA}(R)$ at $R = R_{DA}$ and $R = R_{AA}$, respectively, is not guaranteed. For these reasons, the use of longer buses tends in general to decrease the fidelity and should be limited to what is strictly necessary. However, high-fidelity gates can nevertheless be obtained in many cases: Figure~\hyperref[fig:gate_surface_code]{\ref{fig:gate_surface_code}(c)} shows the time-optimal phases $\phi_{D,A}(t)$ that implement the transversal CZ gate between surface codes of distance $d=11$, using two ancilla atoms in the bus and with lattice spacing $r_0=2.62\,\mathrm{\mu m}$, principal quantum numbers $n_{\rm Rb}=73$ and $n_{\rm Cs}=76$. The pulse duration is $ T_{\rm gate}^{\rm(2a)}\Omega_{\rm max}=10.3$ and for $\Omega_{\rm max}=2\pi\times 2\,\mathrm{MHz}$ the simulated infidelity is $1-F=5.3\times10^{-3}$, dominated by spontaneous emission.

In App.~\ref{app:scalability}, we show that even for $n_A=2$ the gate duration $T^{\rm(2a)}$ and the time spent in the Rydberg manifold $\tau_{\rm ryd}^{\rm(2a)}$ exhibit the same functional scaling as in Eqs.~\eqref{eq:T_gate_1a} and \eqref{eq:tau_ryd_1a}, up to different numerical coefficients. This suggests that, also for atomic buses with multiple ancilla mediators, both the gate fidelity and the achievable range can be enhanced by reducing the lattice spacing $r_0$ and increasing the principal quantum number $n$ and the Rabi frequency $\Omega_{\rm max}$. The resulting longer-range gates could support transversal operations in surface codes of larger distance $d>25$, as well as transversal operations between non-nearest-neighbor logical patches in smaller-distance surface codes.

\subsection{Bell-pair protocol, gate scheduling, and correlated decoding \label{sec:bellpair}}

To synthesize the full logical Clifford group, a fault-tolerant architecture must support a multi-qubit logical entangling operation, typically the controlled-NOT (CNOT) gate, along with the single-qubit Hadamard (H) and phase (S) gates. Within the CSS formalism, the logical CNOT can be executed transversally by applying physical CNOT gates pairwise between corresponding data qubits of two distinct code blocks. In planar two-dimensional architectures, however, executing this bitwise operation between non-overlapping code patches necessitates long-range physical connectivity scaling as $\mathcal{O}(d)$ with code distance $d$.

In this section, we apply our static architecture to investigate a fault-tolerant logical Bell-pair generation protocol between surface codes. This scheme features the execution of a transversal logical CNOT (tCNOT) gate between two distant surface code patches, followed by joint measurement of the resulting logical state. Even though neutral-atom platforms natively support CZ gates rather than CNOT gates, one can decompose the transversal logical operation into CZ and Hadamard gates.

We consider distance-$d$ rotated surface-codes comprising $d^2$ data qubits and $d^2- 1$ ancilla qubits, each code encoding a single logical qubit. A logical tCNOT between two surface codes is implemented by applying $d^2$ physical CNOT gates pairwise between corresponding data qubits. Because syndrome extraction is needed only before and after—not during—the entangling operation, the same ancilla qubits readily available in the code arrays can serve as atomic bus to mediate long-range entanglement.

To maximize parallelism while limiting crosstalk, physical CNOTs mediated by a single ancilla atom can be executed in parallel on atoms that are more than $\lceil d/2\rceil$ rows apart, while they need to be executed sequentially across the columns, requiring a total of $(\lceil d/2\rceil+1) \times d$ time steps on average.  For $d=9$, the larger surface code distance mediated by a single ancilla atom discussed in this work, we assume a Rabi frequency of $\Omega=2\pi\times2$ MHz, an illumination time of $\sim0.65\,\mu$s per gate, and a switching time of $\sim1\,\mu$s between gate layers. Neglecting illumination and switching times for the single-qubit Hadamard gates (which can be implemented globally in only two steps) and the resetting time for the ancilla qubits, this amounts to a total of $88\, \mu$s to execute a single transversal CZ statically, which is far below the ms timescales of shuttling-based architectures. The scheme's parallelism could be improved by using more mediator atoms in the chain at the price of longer Rydberg times and, consequently, lower gate fidelities, or with comparable times and fidelities by increasing the laser Rabi frequency and reducing the lattice spacing. The same ancillas can be reused to entangle different pairs of data qubits. However, to preserve fault tolerance and prevent error propagation, those ancillas must be reset between uses, which in a dual‑species platform can be achieved statically via fast separate controls on the ancilla species.
  
Recent theoretical work \cite{Zhou_2025,sk5y-25b1,cain2025fastcorrelateddecodingtransversal} has shown that joint correlated decoding of logical observables reduces the number of required syndrome extraction rounds per transversal gate from $\mathcal{O}(d)$ to $\mathcal{O}(1)$, at the expense of increased classical decoding complexity. This can be achieved by decoding only the stabilizers across the back-propagation path of \emph{reliable} (opposed to \emph{fragile}) logical Pauli products, i.e. products of logical operators which can be back-propagated through the circuit and terminated at logical resets in the same basis \cite{sk5y-25b1,cain2025fastcorrelateddecodingtransversal}. 

\begin{figure}[t!]
    \centering
    \includegraphics[width=0.9\linewidth]{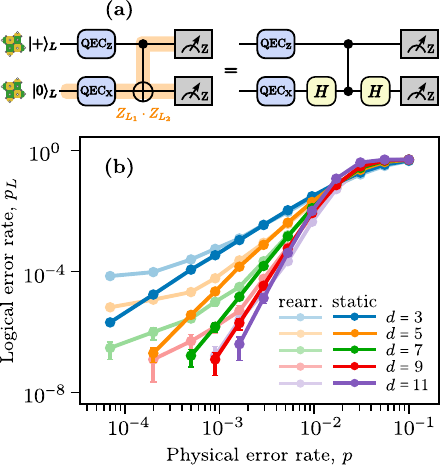}
    \caption{(a) Circuit for the Bell-pair protocol. In orange we highlight the back-propagation path of the reliable logical Pauli product $Z_{L_1}\cdot Z_{L_2}$. (b) Logical error rate as a function of physical error rate for the Bell-pair protocol. Darker lines instead denote simulations for a static architecture using long-range entanglement via Rydberg atomic buses. Lighter lines denote simulations for an architecture using long-range entanglement via atom-rearrangement and suffering from motion-induced errors by heating. Error bars denote standard deviations on Monte Carlo sampling.}
    \label{fig:tcnot}
\end{figure}

In this work, we adopt an analogous correlated decoding strategy, and use the Clifford simulator \texttt{Stim} \cite{Gidney_2021} to simulate the logical Bell-pair protocol as follows. We first prepare the control code in the logical $\ket{+}_L$ state and the target code in the logical $\ket{0}_L$ state, apply the tCNOT gate, and finally measure both codes in the $Z$ basis to read out their logical states [Fig.~\hyperref[fig:tcnot]{\ref{fig:tcnot}(a)}]. The logical $\ket{0}_L$ ($\ket{+}_L$) state is the $+1$-eigenstate of the $Z_L$ ($X_L$) logical operator and is prepared by initializing the qubit register to $\ket{0}^{\otimes d^2}$ ($\ket{+}^{\otimes d^2}$) and measuring only the $X$ ($Z$) stabilizers. After the logical tCNOT, we measure all the data qubits in the $Z$-basis, such that, in total, we perform a single syndrome extraction round in each surface code. 

The logical operators of control and target surface codes, $Z_{L_1}$ and $Z_{L_2}$ respectively, are individually fragile, but correlated through their product $Z_{L_1}\cdot Z_{L_2}$, which is instead reliable. We decode only the stabilizers in the back-propagation path of $Z_{L_1}\cdot Z_{L_2}$ [in orange in Fig.~\hyperref[fig:tcnot]{\ref{fig:tcnot}(a)}], which is equivalent to correct only $X$ errors, since $Z$ errors cannot affect the final logical measurement in the $Z$-basis. In \texttt{Stim}, we use \texttt{detector} annotations, defined as the binary addition of measurement outcomes, to keep track of errors. Here, we only define \texttt{detectors} at the end of the circuit to reconstruct $Z$ stabilizers of control and target codes from the final logical measurement. For the target code, these \texttt{detectors} additionally compare the sign of the reconstructed target $Z$ stabilizers with the measurement outcomes of the corresponding $Z$ ancilla qubits of the control code before the tCNOT, as $X$ errors on the control can spread $X$ errors onto the target. We note that $Z$ and $X$ stabilizer measurements of the initial syndrome extraction round in the control and target codes are 50-50 random, therefore we do not define any \texttt{detectors} before the tCNOT. 

\subsection{Noise model and numerical simulations \label{sec:sc-noise}}
In the simulations, we use a noise model with error strengths set according to the typical operation infidelities measured in neutral atom experiments. All CNOT gates are decomposed into CZ and Hadamard gates. We assume that the CZ gates used for syndrome extraction experience two-qubit depolarizing noise with probability $p$, while single-qubit gates are assumed noiseless. Measurement and reset in the $Z$ basis are injected with bit-flip errors with probability $p$. Idle errors are negligible in static settings: While the static paradigm features reduced parallelism, local addressing ensures that atoms not involved in long-range gates experience no pulses and remain stationary. Additionally, crosstalk-induced errors can be made negligible via appropriate adjustment of the lattice spacing or through pulse-level optimization \cite{Warttmann_2026}. Ancilla qubits used in the bus to mediate the long-range CZ gates are measured before the tCNOT and reset after. We simulate the noise on the atomic bus-mediated long-range gates in the tCNOT as range-dependent two-qubit depolarizing errors.

We calculate the scaling of the long-range gate infidelity as a function of the nearest-neighbor gate infidelity (time-optimal pulse assumed) accounting for errors due to spontaneous emission, motion and blockade imperfection as discussed in Sec.~\ref{sec.operations_surface_code}, and accordingly inject depolarizing errors with probability $p^*=f(d)\cdot p$, where $f(d)$ is a code distance-dependent function. \rsub{Table~\ref{tab:architecture_errors} summarizes the error rates of the noise model used for the numerical simulations of a surface code of distance $d$ in the static and rearrangement architectures.}

\begin{table}[t!]
    \centering
    {\color{black}
    \begin{tabular}{c|c|cc}
        \toprule
        operation & noise & static  &  rearrangement \\
        \hline\hline
        nearest-neighbor CZ & \texttt{DEPOLARIZE2}  & $p$     & $p$ \\
        $Z$-measurement & \texttt{X}$\_$\texttt{ERROR}  & $p$ & $p$ \\
        $Z$-reset & \texttt{X}$\_$\texttt{ERROR}  & $p$ & $p$ \\
        Hadamard gate & \texttt{DEPOLARIZE1}   & 0 & 0 \\
        long-range CZ & \texttt{DEPOLARIZE2}  & $p\cdot f(d)$ & $p$ \\
        atomic shuttling & \texttt{DEPOLARIZE1}  & 0 & $p_{\rm idle}=A\cdot d^{2/3}$\\
        \hline
        decoder & & BPOSD & BPOSD \\
        \bottomrule
    \end{tabular}
    }
    \caption{\rsub{Total error probabilities for each noise channel used in \texttt{Stim} for the numerical simulations of distance-$d$ surface codes in the static and rearrangement-based settings. Long-range CZ gates are modeled differently in the static and rearrangement-based settings. Static: range-dependent gate errors $p\cdot f(d)$. Rearrangement: two parallel moves (out and back) add two idle layers at rate $p_{\rm idle}=A\cdot d^{2/3}$ ($A=6\times10^{-4}$, see App.~\ref{app.heating}) and all gates are nearest-neighbor with error rate $p$. Movements between interaction and measurement zones (typical of zoned architectures) are not simulated. 
    }}
    \label{tab:architecture_errors}
\end{table}

Recall that the infidelity of the short-range time-optimal gate is fundamentally limited by spontaneous emission, $1-F^{\rm(to)}=\Gamma \tau_{\rm ryd}^{\rm(to)}\approx3\Gamma/\Omega_{\rm max}$ \cite{Jandura2022}. Spontaneous emission is also the main limitation of symmetric one-mediator bus gates, whose time spent in the Rydberg manifold is in Eq.~\eqref{eq:tau_ryd_1a}. Therefore, we take for $f(d)$
\begin{equation}
    \label{eq:f}
    f = \frac{1-F^{\rm (1a)}}{1-F^{\rm (to)}} \approx 1.7 + \frac{1.1}{(B_{DA}/\Omega_{\rm max})^{2}}.
\end{equation}
For code distances $d=3-9$ we can find appropriate $n_{\rm Rb,Cs}<80$ such that $B_{DA}/\Omega_{\rm max}\approx2$, thus $f(d)\approx2$. For $d=11$, we need to consider two ancilla atoms in the bus, and $f(d)\approx3.7$ [see App.~\ref{app:scalability}]. Larger distances require larger numbers of ancilla mediators or larger principal quantum numbers $n$ and are not simulated.

\rsub{The time-optimal phase profile of the bus gates is smooth and continuous, but typically slightly more complex than the simple sinusoidal shape of short-range perfect-blockade gates \cite{Jandura2022}, potentially increasing their sensitivity to amplitude, phase, or calibration errors. However, since this sensitivity depends strongly on the specific experimental setup and does not represent a fundamental limitation of our approach, here we assume, for simplicity, a similar level of control over the pulse shaping for both short-range perfect-blockade and long-range bus gates.}

We compare this stationary paradigm with a method employing atom rearrangement to realize the tCNOT across distant code patches. Single-qubit gate errors are again neglected, as they are typically an order of magnitude smaller than two-qubit gate and measurement errors. Moreover, since these errors impact both paradigms equally, their contribution can be rescaled. We assume that \emph{all} CZ gates experience two-qubit depolarizing errors with probability $p$, because CZ gates used for the tCNOT are local upon atom shuttling in this setting. We inject bit-flip errors with probability $p$ on both measurement and reset gates in the $Z$ basis as in the stationary setting.

An additional error source that must be considered in atom-shuttling architectures is dephasing induced by motional heating. To execute a tCNOT gate between two surface codes, one patch is shuttled over the other, nearest-neighbor entangling gates are executed using global laser pulses, and the mobile patch is subsequently transported back. Assuming two adjacent distance-$d$ surface code patches, the movement cost for a tCNOT is $\mathcal{O}(d)$. Using the minimum-jerk movement trajectory demonstrated in Ref.~\cite{Finkelstein2024,Chinnarasu2025,Rines2025}, the wall-clock time of such movement operations scales as $\mathcal{O}(d^{1/3})$, and consequently the infidelity as $\mathcal{O}(d^{2/3})$ (see  Appendix~\ref{app.heating}). In our simulations, we model these errors as single-qubit depolarizing errors affecting only the moving code patch—conventionally, the target one—at two time steps, before and after the tCNOT. We choose depolarizing—and not phase-flip errors—to qualitatively account for possible shuttling-induced leakages out of the computational subspace. We assume an error strength of $p_\text{idle}=A\cdot d^{2/3}$, where $A$ is a constant describing the error rate for shuttling one atomic qubit over one(settice spacing ($d=1$) [see App.~\ref{app.heating}].
\rsub{There have been several recent measurements of atom loss and infidelity due to tweezer transport with alkali atom hyperfine qubits \cite{Manetsch2025,JCWang2026}. Measurements were made with different distances, trap depths, and velocity profiles, from which it is difficult to extract a single value of $A$.  Here we make an optimistic estimate of $A=6\times 10^{-4}$. }
  We emphasize that $p_\text{idle}$ is assumed to be independent of $p$. This assumption is motivated by the fact that while active gate errors can theoretically be mitigated through increased laser power, heating effects impose an intrinsic architecture-dependent bottleneck that must be addressed via separate strategies.

\rsub{
We note that, while the hyperfine encoding used in this work is subject to dephasing due to the differential polarizability of the qubit states, this is not the case for nuclear-spin qubits, e.g., in metastable alkaline-earth(-like) atoms~\cite{Zhang2026}. However, the latter remain subject to other transport-related error sources, such as atom loss and leakage, and transport errors can therefore remain significant [see App.~\ref{app.heating}].}

We decode syndrome information with Belief-Propagation with Ordered-Statistics Decoder (BPOSD) \cite{Roffe_2020}, which is capable of handling the \emph{hyperedges} in the decoding graph arising from errors propagating through transversal operations. We use Belief Propagation in \texttt{min-sum} mode with scaling factor $s=0.3$ (empirically optimized to yield the lowest possible logical error rate below threshold). Ordered Statistics Decoder is used in combination sweep mode, $\texttt{osd-cs}$, at first order.

We show the results of Monte Carlo simulations for the Bell-pair protocol in Fig.~\hyperref[fig:tcnot]{\ref{fig:tcnot}(b)}. We plot logical error rate, calculated as the ratio between numbers of decoder failures and Monte Carlo samples, as a function of the physical error rate, $p$. In Fig.~\hyperref[fig:tcnot]{\ref{fig:tcnot}(b)}, darker lines denote the static, atom-bus mediated protocol, while lighter lines denote the one using atom rearrangement. We highlight that the horizontal axis in this plot corresponds to the physical error rate of nearest-neighbor gates, thereby representing the infidelity of the ``best gates" within the static architecture. Long-range gates are instead accounted to be lower fidelity and are accordingly injected with higher noise rates as described above.
The results shown in Fig.~\hyperref[fig:tcnot]{\ref{fig:tcnot}(b)} reveal that the atom rearrangement-based protocol is completely bottlenecked by heating errors, which are responsible for the error floor visible at low $p$ for the lighter lines. Figure~\hyperref[fig:tcnot]{\ref{fig:tcnot}(b)} also shows that near the threshold (crossing point at $p\sim10^{-2}$), gate errors outweigh heating effects, resulting in the static architecture performing slightly worse than the rearrangement-based scheme (e.g., the $d=11$ violet curve). This is, however, no longer the case in the low-$p$ regime, where heating dominates over gate errors. Notably, for all the code distances displayed, the static architecture shows about one order of magnitude gain over the atom-shuttling approach at $p\sim10^{-3}$, corresponding to fidelities within reach of current neutral atom quantum computers. 

These findings establish a viable pathway toward efficient, static, and fault-tolerant long-range entanglement between logical qubits. Furthermore, we note that, while our discussion focuses on surface codes, the very same protocol seamlessly extends to any CSS code, including high-rate CSS quantum LDPC code families \cite{Bravyi_2024,Pecorari_2025,PhysRevA.111.022433}. Such extensions would enable higher-rate Bell-pair generation and, more broadly, the execution of logical transversal operations between high-rate codes~\cite{pecorari2025addressablegatebasedlogicalcomputation}.

\section{Long-range stabilizer readout\label{sec.long_range_stabilizer_readout}}

Another key building block for a FTQC architecture is a flexible stabilizer readout scheme that is capable of accommodating QEC codes with long-range stabilizer gates, such as high-rate quantum LDPC codes. These codes offer higher encoding rates compared to the surface code, which makes them highly efficient quantum memories. \rsub{On the other hand} the computing is generally hard, requiring extra resources (e.g., auxiliary physical qubits), or switching to lower-rate codes, such as surface codes, where the full logical Clifford gate set can easily be implemented transversally.

In this section, we design a novel stabilizer readout scheme for two families of quantum LDPC codes, namely La-cross and bivariate bicycle, where the codes' long-range connectivity is achieved statically using mediator atoms. In Sec.~\ref{sec.gate_lacross} we design the time-optimal and motion-robust gate pulses to implement long-range stabilizer gates, which, contrary to the previous section, now must couple data and ancilla qubits, resulting in intraspecies interactions. In Sec.~\ref{sec.lacross} we first apply our machinery to La-cross codes, discussing gate robustness and performing large-scale numerical simulations. We again compare the performance of the static architecture against one using atom shuttling to target the same long-range connectivity, and discuss gate scheduling and parallelism of our scheme. Sec.~\ref{sec.bb}, instead, extend the framework of long-range atom-bus gates to bivariate bicycle codes. We discuss code layout and gate scheduling for these codes as well.

\begin{figure*}[t!]
    \centering
    \includegraphics[width=1\linewidth]{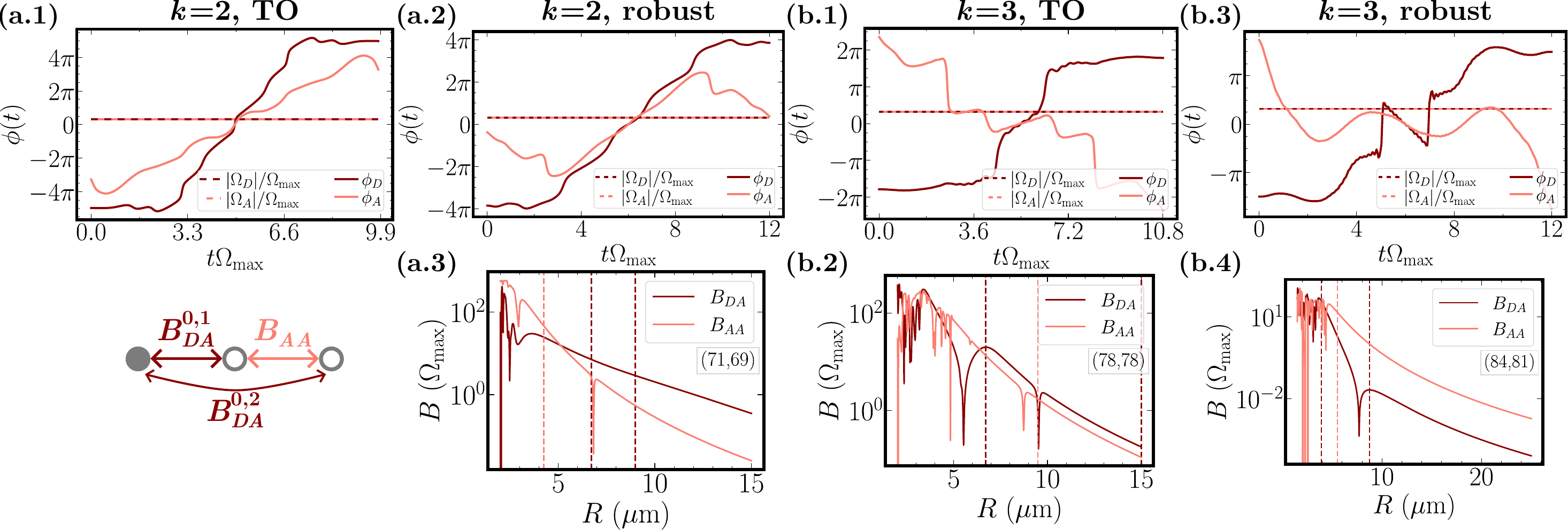}
    \caption{CZ gates for long-range stabilizers of the La-cross code. \textbf{(a)} Gates for a $k=2$-La-cross code. Panel (a.1) shows the phases $\phi_{D,A}$ and amplitudes $|\Omega_{D,A}|$ of the time-optimal pulse, while panel (a.2) shows the corresponding motion-robust pulse. The associated data--ancilla (dark line) and ancilla--ancilla (light line) interaction potentials are shown in panel (a.3). The calculations use $(n_{\rm Rb}=71,n_{\rm Cs}=69)$, lattice spacing $r_0=3\,\mathrm{\mu m}$, and Rabi frequency $\Omega_{\rm max}=2\pi\times2\,\mathrm{MHz}$.\textbf{(b)} Gates for a $k=3$-La-cross code. Panels (b.1) and (b.2) show, respectively, the time-optimal pulse and the corresponding interaction potentials for $(n_{\rm Rb}=78,n_{\rm Cs}=78)$, $r_0=3\,\mathrm{\mu m}$, and $\Omega_{\rm max}=2\pi\times2\,\mathrm{MHz}$. Panels (b.3) and (b.4) show an enhanced-parameter implementation and its corresponding interaction potentials for $(n_{\rm Rb}=84,n_{\rm Cs}=81)$, $r_0=1.75\,\mathrm{\mu m}$, and $\Omega_{\rm max}=2\pi\times8\,\mathrm{MHz}$. In all potential plots, the dark and light curves correspond to the data--ancilla and ancilla--ancilla interactions, respectively. In all potential plots the vertical dashed lines represent the interatomic distances $R_{AA}$, $R_{DA}^{0,1}$ and $R_{DA}^{0,2}$.}
    \label{fig:lacross_gate}
\end{figure*}

\subsection{Asymmetric atomic bus gates for qLDPC codes \label{sec.gate_lacross}}

The long-range stabilizer operators of qLDPC codes, such as the $k$-La-cross codes or the bivariate bicycle code, require the implementation of a CZ gate between a data atom and an ancilla atom that are not nearest neighbors [configuration (ii)]. To mediate the interaction, an additional ancilla atom that does not lie in the same row or column as the other two atoms is introduced. \rsub{In order to perform in-place, global readout of the ancilla qubits only (as required between quantum-error-correction rounds) data and ancilla atoms must be of different atomic species. Therefore, the presence of two consecutive atoms of the same species in the three-atom bus makes it inherently asymmetric}, and the interaction Hamiltonian in Eq.~\eqref{eq:interaction} takes the form
\begin{align}
    \nonumber
    H_{\rm int}^{\rm (1b)} = {} & B_{DA}^{0,1} \ket{r_D r_A}\bra{r_D r_A}_{0,1}
    + B_{DA}^{0,2} \ket{r_D r_A}\bra{r_D r_A}_{0,2} \\
    \label{eq:interaction_1b}
    & + B_{AA}^{1,2} \ket{r_A r_A}\bra{r_A r_A}_{1,2},
\end{align}
where the label $(\rm 1b)$ refers to an asymmetric configuration with only one mediator: data-ancilla-ancilla. Here $B_{DA}^{i,j}$ denotes the data--ancilla (interspecies) vdW interaction between atoms $i$ and $j$, separated by a distance $R_{DA}^{i,j}$, while $B_{AA}^{1,2}$ denotes the ancilla--ancilla (intraspecies) vdW interaction between ancilla atoms $1$ and $2$. The total Hamiltonian reads
\begin{equation}
    \label{eq:H_tot_1b}
    H(t) = H_{D}^{(0)}(t) + H_A^{(1)}(t) + H_{A}^{(2)}(t) + H_{\rm int}^{\rm (1b)}.
\end{equation}

We show in the subsections below via numerical optimization that this asymmetry, together with the additional interaction energy scales, leads to somewhat longer pulses than for symmetric pulses. Moreover, compared to the symmetric pulses of configuration (i), we find a stronger sensitivity to motion. This effect has at least two causes: (a) the atoms are closer, and thus the relative displacement due to thermal motion is larger; (b) given the asymmetric ancilla–ancilla–data bus, it is impossible to choose a lattice spacing $r_0$ and principal quantum numbers $n_{\rm Cs}$ and $n_{\rm Rb}$ such that sufficiently strong blockades are obtained while simultaneously operating near a local maximum of both the interaction potentials $B_{DA}(R)$ and $B_{AA}(R)$. Despite this, efficient stabilizer readout schemes can be obtained, as we show below. 

\subsection{La-cross codes \label{sec.lacross}}

\subsubsection{Gates performance and robustness}

Figure~\hyperref[fig:lacross_gate]{\ref{fig:lacross_gate}(a.1)} (\hyperref[fig:lacross_gate]{\ref{fig:lacross_gate}(b.1)}) shows the time-optimal phases $\phi_{D,A}(t)$ that implement the long-range CZ gate for the stabilizers in a $k=2\,(k=3)$-La-cross code, using a single ancilla bus and with lattice spacing $r_0=3\,\mathrm{\mu m}$, principal quantum numbers $n_{\rm Rb}=71\,(78)$ and $n_{\rm Cs}=69\,(78)$. We find a pulse duration $T^{\rm(1b)}\Omega_{\rm max}=9.82\,(10.8)$ and a simulated infidelity $1-F=1.6\,(1.0)\times10^{-2}$, dominated by both thermal motion and spontaneous emission. We note that while these individual infidelities exceed the error-correction threshold for La-cross codes, long-range gates are executed less frequently than nearest-neighbor ones. Specifically, only two long-range gates are required by each stabilizer, and sometimes only one by some boundary stabilizers. Consequently, the impact of these operations can be effectively compensated for by the higher fidelity of the local gates, allowing the total error rate to remain below the effective fault-tolerance threshold. We provide quantitative evidence for this behavior in the following sections through large-scale numerical simulations of La-cross codes under realistic noise models.

We find that the sensitivity to variations in the blockade potential can be reduced, for example, by optimizing the gates not just for a single value of the blockade strength $B_{DA}$, but over a set of values in a neighborhood of the expected value $B_{DA}$ [see Sec.~\ref{sec.gate_1ancilla}], at the cost of longer pulses and therefore increased infidelity due to spontaneous emission. For example, applying this further robust-optimization the infidelity of the pulse for $k=2$ [see Fig.~\hyperref[fig:lacross_gate]{\ref{fig:lacross_gate}(a.2)}] is reduced by a factor 2 as $1-F=7.4\times10^{-3}$, now limited by spontaneous emission only.

The gate infidelity can be further reduced by employing higher principal quantum numbers, stronger laser driving, and smaller lattice spacings, provided that laser intensity and phase noise remain under control and that external-field fluctuations do not significantly perturb the Rydberg states. For instance, in the case of a $k=3$-stabilizer, using $(n_{\rm Rb}=84,n_{\rm Cs}=81)$, $\Omega_{\rm max}=2\pi\times8\,\mathrm{MHz}$, and $r_0=1.75\,\mathrm{\mu m}$, we obtain an optimized pulse \rsub{[see Fig.~\hyperref[fig:lacross_gate]{\ref{fig:lacross_gate}(b.3)}]} with infidelity $1-F=7.1\times10^{-4}$. In this last case, motion-robustness is achieved by choosing an interaction potential that provides an almost perfect blockade between nearest-neighbor atoms, $B_{DA}^{0,1},B_{AA}\gg\Omega_{\rm max}$, while exhibiting a local maximum at the next-nearest-neighbor data--ancilla distance, $B_{DA}^{0,2}\ll\Omega_{\rm max}$ \rsub{[see Fig.~\hyperref[fig:lacross_gate]{\ref{fig:lacross_gate}(b.4)}]}, thereby substantially reducing its sensitivity to variations in those parameters. \rsub{A complete list of the pulses optimized for long-range stabilizer readout in La-cross code, together with their parameters, is provided in Table~\ref{tab:all_lacross_gates}.} 

\subsubsection{Noise model and numerical simulations}

With \texttt{Stim}, we perform memory simulations of $k=2$- and $k=3$-La-cross codes encoding $4$ and $9$ logical qubits, respectively. These code instances feature moderate long-range connectivity and are hence highly suitable for static implementations. We simulate $d$ rounds of syndrome extraction to ensure robustness against measurement errors and again employ a noise model with parameter sets specific to neutral atom platforms. We assume single-qubit gates to be noiseless, while measurements and resets in the $Z$($X$)-basis are injected with bit-flip (phase-flip) errors with probability $p$. Two-qubit gates experience range-dependent depolarizing errors to mimic a static implementation with atom-bus long-range entanglement. The strength of these errors is calculated from microscopic gate modeling, assuming both time-optimal and robust pulses, as follows. For $k=2,3$-La-cross codes, we consider atomic busses made of a single ancilla atom to maximize gate fidelity, since syndrome extraction gates act over moderately long distances $r$. We evaluate the scaling of long-range gate infidelity as a function of nearest neighbor (time optimal pulse assumed) gate infidelity and accordingly inject depolarizing errors with probability $p^*=f(r)\cdot p$ as described in Sec.~\ref{sec.gate_lacross} [see Eq.~\eqref{eq:f}]. The values of $f(r)$ used in the simulations are $f(r)\approx6$ for time-optimal gates and $f(r)\approx2$ for robust gates. In the simulation of this static configuration, idle errors are neglected, which can be accomplished by optimizing the lattice spacing or applying pulse-level mitigation techniques \cite{Warttmann_2026}. For $k>3$-La-cross codes, larger numbers of mediator ancilla qubits are necessary, and are not simulated. 

We compare this static setting with one that uses parallel atom rearrangement. Since La-cross codes have open boundary conditions, the total connectivity required for their implementation does not scale with the system size and only depends on $k$. Additionally, because $X$ and $Z$ stabilizers have two long-range gates, each aligned along the same direction, La-cross codes can be implemented using only four parallel long-range moves (two moves per stabilizer type). We assume that the ancilla blocks move towards the targeted data qubits. Thus, we simulate the motion-induced heating errors affecting these long-range transports as single-qubit depolarizing noise only on the ancilla qubits involved in the move (idle errors). Each ancilla block needs to be moved twice (in and back), therefore we model a total of four layers of errors. Idle errors affecting spectator qubits are assumed to be negligible since these atom do not move. The strength of the idle errors impacting the ancilla qubits is modeled in the same way as we did above for the surface code following Ref.~\cite{Rines2025}. That is, we assume an error strength of $p_\text{idle}=A\cdot R^{2/3}$, where $R$ is the gate extent in units of lattice spacing and $A$ is a constant describing the error rate for shuttling one atomic qubit over one lattice spacing ($R=1$), which we assume to be
$6 \times 10^{-4}$ based on optimistic estimates from current experiments (approximately corresponding to a fidelity of
$99.97\%$) \cite{Manetsch2025}. The gate range $R$ solely depends on $k$ as $R=2k-1$. We again note that, as for the Bell-pair protocol, $p_\text{idle}$ is assumed to be independent of $p$. 

We decode syndrome information with BPOSD: we use Belief Propagation in \texttt{min-sum} mode with scaling factor $s=0.3$. Ordered Statistics Decoder is used in combination sweep mode, \texttt{osd-cs}, at first order.

\begin{figure*}[t!]
    \centering
    \includegraphics[width=.9\linewidth]{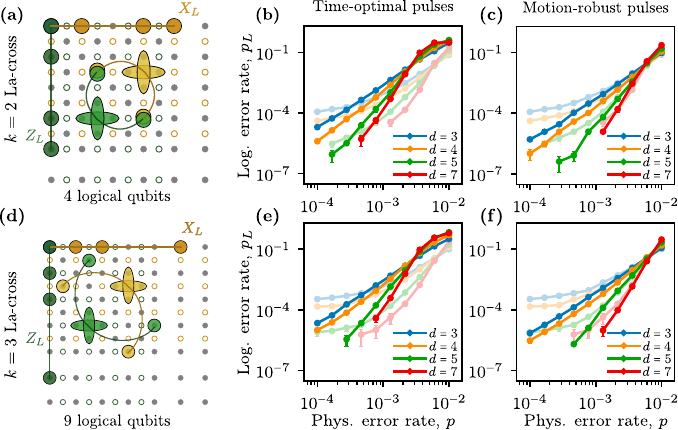}
    \caption{$k$-La-cross code encoding $K=k^2$ logical qubits. Data qubits (filled gray dots) form a $n\times n$ main lattice and a $(n-k)\times(n-k)$ sub-lattice. $X$-type (empty green dots) and $Z$-type (empty yellow dots) ancillas occupy $(n-k)\times n$ and $n\times(n-k)$ rectangular lattices, respectively. Representative $X$ (green) and $Z$ (yellow) stabilizers, alongside $X_L$ and $Z_L$ logical operators, are shown. (a,d) Layouts of $k=2$ and $k=3$ La-cross codes. (b,c) Quantum memory simulations for $k=2$ La-cross codes with noise models based on atom rearrangement (light), and static arrays using either \rsub{time-optimal} (b) or \rsub{motion-robust} (c) pulses. (e,f) Quantum memory simulations for $k=3$ La-cross codes with noise models based on atom rearrangement (light), and static arrays using either \rsub{time-optimal} (e) or \rsub{robust} (f) pulses. Error bars correspond to standard deviations on Monte Carlo samplings. See main text for the simulated code parameters.}
    \label{fig:lacross-memory}
\end{figure*}

We show the results of Monte Carlo simulations for $k=2$ [Fig.~\hyperref[fig:lacross-memory]{\ref{fig:lacross-memory}(a)}] in panels~\hyperref[fig:lacross-memory]{(b)} and ~\hyperref[fig:lacross-memory]{(c)} of Fig.~\hyperref[fig:lacross-memory]{\ref{fig:lacross-memory}} assuming \rsub{time-optimal and motion-robust pulses}, respectively. Numerical results assuming \rsub{time-optimal and motion-robust pulses} for $k=3$-La-cross codes [Fig.~\hyperref[fig:lacross-memory]{\ref{fig:lacross-memory}(d)}] are instead shown in panels~\hyperref[fig:lacross-memory]{(e)} and~\hyperref[fig:lacross-memory]{(f)} of Fig.~\hyperref[fig:lacross-memory]{\ref{fig:lacross-memory}}, respectively. We plot the per-round logical error rate, normalized as $p_L\rightarrow1-(1-p_L)^{1/d}$, as a function of the physical error rate $p$ affecting the ``best" nearest-neighbor gates. Dark decoding curves simulate a static implementation via long-range atom-bus gates with optimized gate pulses, whereas light decoding curves simulate an implementation via atom rearrangement. For both $k=2,3$ code families, we simulate code distances $d=3,4,5,7$, corresponding to the codes $[[34,4,3]]$, $[[52,4,4]]$, $[[100,4,5]]$, $[[202,4,7]]$ ($k=2$) and $[[45,9,3]]$, $[[65,9,4]]$, $[[149,9,5]]$, $[[269,9,7]]$ ($k=3$). 

\rsub{The numerical simulations in Fig.~\hyperref[fig:lacross-memory]{\ref{fig:lacross-memory}(b,c,e,f)} compare the static implementation of $k=2,3$ La-cross codes against the rearrangement-based one, considering two settings for the static case: time-optimal and motion-robust gate pulses. In both settings the static implementation achieves significant reductions in logical error rate over the rearrangement-based one, though in different regimes. With time-optimal pulses [Fig.~\hyperref[fig:lacross-memory]{\ref{fig:lacross-memory}(b,e)}], the static implementation has an error-correction threshold of $\sim3\times10^{-3}$, approximately a factor of $2$ below that of the rearrangement-based implementation ($\sim6\times10^{-3}$). It therefore shows higher logical error rates for $p\gtrsim4\times10^{-4}$ but outperforms the rearrangement-based scheme for $p\lesssim4\times10^{-4}$, where the latter is limited by its heating-induced error floor. With motion-robust pulses [Fig.~\hyperref[fig:lacross-memory]{\ref{fig:lacross-memory}(c,f)}], the results improve substantially: the static threshold becomes comparable to the rearrangement-based one, around $6\times10^{-3}$ for both La-cross code instances. The crossover also shifts upward, so that the static architecture already shows lower logical error rates for $p\lesssim2\times10^{-3}$. For code distances $d=3-7$ and nearest-neighbor gate error rate $p\approx4\times10^{-4}$, it achieves more than an order of magnitude improvement over the rearrangement-based architecture. We observe that the crossover error rate of $p\approx2\times10^{-3}$ corresponds to a nearest-neighbor  two-qubit gate infidelity that lies within reach of recent neutral-atom experimental demonstrations~\cite{Evered2026, Liu2026}.} These numerical results therefore indicate static atom-bus long-range entanglement with motion-robust gate pulses as an efficient pathway toward the fast and static implementation of high-rate quantum LDPC codes.

\subsubsection{La-cross gate scheduling}
\begin{figure}[t!]
    \centering
    \includegraphics[width=\linewidth]{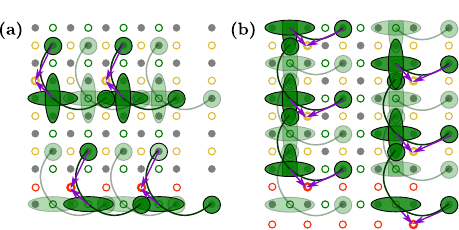}
    \caption{Gate scheduling for $Z$ stabilizers of La-cross codes using a single $X$ ancilla atom as mediator ($k=2$ assumed in the example). An analogous scheduling holds for $X$ stabilizers. (a) Scheduling of long-range \emph{north} gates. (b) Scheduling of long-range \emph{east} gates. Gates with same gradient of color can be performed in parallel (see main text). At the boundaries of the qubit arrays, extra atoms \rsub{(in red)} must be added to mediate long-range entanglement.}
    \label{fig:lacross-schedu}
\end{figure}

We now discuss optimal gate scheduling of La-cross codes when implemented statically via Rydberg atomic busses made of a single ancilla atom. First, we note that, contrary to the Bell-pair protocol between surface codes, syndrome‑extraction two‑qubit gates act over moderate ranges and entangle ancilla and data qubits, making them inter‑species in dual‑species implementations. We also observe that ancillas cannot be arbitrarily reset during a memory experiment because they carry syndrome information that must not be lost before processing it. We avoid information loss by measuring $X$ and $Z$ stabilizers at separate times, using $Z$ ($X$) ancillas to mediate $X$ ($Z$) long-range stabilizer gates and resetting those ancillas after use. This can be easily achieved in La-cross codes as $X$ and $Z$ ancilla qubits belong to two different lattices. In particular, $X$-type ancilla qubits form a rectangular $(n-k)\times n$ lattice, while the $Z$-type ones form a rectangular $n\times(n-k)$ lattice. To visualize these distinct lattices, we color the $X$ and $Z$ ancilla qubits in green and yellow, respectively, in Fig.~\ref{fig:lacross-schedu}.

Another constraint on optimal gate scheduling is crosstalk: parallel executed gates must act on atoms separated by approximately four lattice spacings. We show the resulting scheduling of $Z$ stabilizers in panels~\hyperref[fig:lacross-schedu]{(a)} and~\hyperref[fig:lacross-schedu]{(b)} of Fig.~\hyperref[fig:lacross-schedu]{\ref{fig:lacross-schedu}} for long-range north and east gates, respectively. Analogous gate scheduling symmetrically applies to $X$ stabilizers as well. North long-range gates can be implemented in parallel every $k+1$ rows (starting from the $(k+1)$th row to avoid boundary stabilizers) and on alternating columns of $Z$ ancilla qubits, as drawn in Fig.~\hyperref[fig:lacross-schedu]{\ref{fig:lacross-schedu}(a)}. One extra row of ancilla atoms must be added at the bottom boundary of the array to mediate long-range entanglement [\rsub{red} empty dots in Fig.~\hyperref[fig:lacross-schedu]{\ref{fig:lacross-schedu}(a)}]. East long-range gates can be implemented in parallel on alternating rows and every $k+1$ columns of $Z$ ancilla qubits, as drawn in Fig.~\hyperref[fig:lacross-schedu]{\ref{fig:lacross-schedu}(b)}. Together with the previously added row of ancilla at the bottom boundary of the array, one more row must be added to measure long-range east gates while preserving parallelism. Alternatively, the same row of ancilla atoms can be used twice in a separate time step. An analogous gate scheduling symmetrically applies to long-range south and west gates of $X$ stabilizers as well, doubling the step count. Nearest-neighbor gates can be measured in eight time steps, four for $X$ and four for $Z$ stabilizers. Thus, the total number of steps per syndrome extraction round to compile a $k$-La-cross code reads  
\begin{equation*}
    (2\times4)+2\times[2\times2\times(k+1)]=8k+16
\end{equation*}
or $8k+17$ if the long-range east gates of the bottom row of stabilizers are measured separately without adding one extra row of atoms. We observe that for very small code distances, such as $d=2$ and $d=3$, this scheduling reduces to a (purely or quasi) sequential implementation of long-range stabilizer gates. Nevertheless, the number of required time steps is crucially independent of the code distance. For $d$ rounds of syndrome extraction, $d\times(8k+16)$ time steps are necessary to compile a $k$-La-cross code. 

Considering the optimal pulses derived in the previous section, we can now quantitatively estimate the time of a single syndrome extraction round for the $k=2$- and $k=3$-La-cross codes. For $k=2$-La-cross codes, assuming a Rabi frequency $\Omega_\text{max}=2\pi\times2$ MHz, the duration of a single time-optimal gate between nearest neighbor atoms is $T^{\rm(to)}=7.6/\Omega_\text{max}\approx0.6\,\mu$s. Instead, for long-range gates with time-optimal pulses $T^{\rm(1b)}=9.8/\Omega_\text{max}\approx0.8\,\mu$s, whereas with motion-robust pulses $T^{\rm(1b,rob)}=12/\Omega_\text{max}\approx1\,\mu$s. Thus, the total illumination time for a single QEC cycle, $8\times0.6+(8\cdot2+8)\times T^{\rm(1b)}$, is approximately $24\,\mu$s with time-optimal pulses and $29\,\mu$s with robust pulses. Assuming a switching time of $1~\mu$s between subsequent gate layers and a measurement time of $100\,\mu$s (we assume one step for $X$ ancilla measurements and one for $Z$ ancilla measurements) and neglecting the contributions from the global single-qubit Hadamard gates, the time of a single QEC cycle time for $k=2$-La-cross codes amounts to $255\,\mu$s with time-optimal pulses and to $260\,\mu$s with motion-robust pulses. Thus, for the largest $d=7$ $k=2$-La-cross code discussed in this work, assuming $d=7$ QEC cycles, the total time for a memory experiment is approximately $1.8$ ms. Under the same assumptions and considering time-optimal pulses, for $k=3$-La-cross codes, a single QEC cycle lasts approximately $276\,\mu$s, and a full memory experiment for a $d=7$ code takes approximately $1.9$ ms. 
Instead, the implementation of quantum LDPC codes via atom rearrangement typically occurs over longer timescales with exact numbers strictly depending on the code layout. In Ref.~\cite{QXu2024} it was estimated that the total time per rearrangement step is $t_\text{step}\approx2\tau_t+2t_\text{move}(\ell_\perp)+t_\text{move}(\ell_\parallel)$, where $2\tau_t$ is the total pick up and drop time, $2t_\text{move}(\ell_\perp)$ is the orthogonal travel time for the lateral move and back, and $t_\text{move}(\ell_\parallel)$ is the longitudinal travel time. For $k=2,3$-La-cross codes, $\ell_\parallel=3r_0,5r_0=15,20\,\mu$m, respectively, for a lattice spacing $r_0=5\,\mu$m. Assuming the same numbers as in \cite{QXu2024}, we estimate $t_\text{step}\approx0.20$--$0.25$ ms for $k=2,3$-La-cross codes, respectively. The number of steps per QEC cycle to implement the long-range stabilizer gates is $8$ (move in and move back twice per stabilizer and assuming $X$ and $Z$ stabilizers are measured separately), resulting in a rearrangement time per cycle of $1.6$ ms ($k=2$) and $2$ ms ($k=3$). Thus, including the illumination time, the total QEC cycle time for an implementation via atom rearrangement (neglecting contributions from single-qubit gates) is $\approx1.8$--$2.2$ ms, and the time for a full memory experiment for a distance $d=7$ code (assuming $d$ QEC cycles) is $\approx14$ ms---approximately one order of magnitude larger than the estimated times for the implementation via static atomic buses.

\subsection{Bivariate Bicycle Codes \label{sec.bb}}

\begin{figure*}
    \centering
    \includegraphics[width=0.7\linewidth]{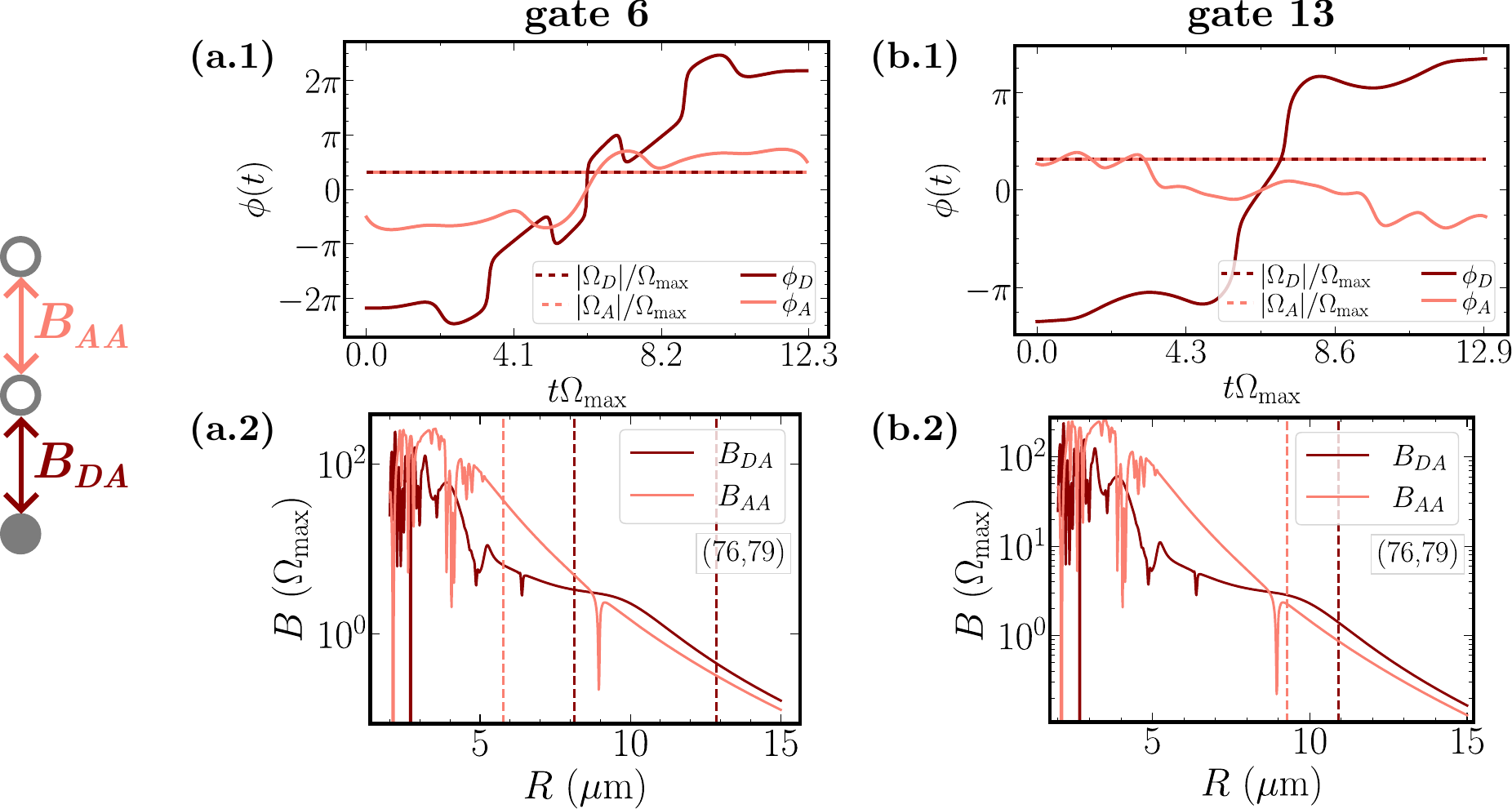}
    \caption{CZ gate for long-range stabilizers of the bivariate bicycle code considering $\Omega_{\rm max}=2\pi\times2\,\mathrm{MHz}$. \textbf{a)} Phases $\phi_{D,A}$ and amplitudes $|\Omega_{D,A}|$ of the time-optimal gate 6 of a BB code, with $B_{DA}^{0,1}/\Omega_{\rm max}=3.31$, $B_{DA}^{0,2}/\Omega_{\rm max}=0.45$ and $B_{AA}/\Omega_{\rm max}=37.6$ (a.1), together with the corresponding data--ancilla (dark line) and ancilla--ancilla (light line) interaction potentials for principal quantum numbers $n_{\rm Rb}=76$ and $n_{\rm Cs}=79$ and lattice spacing $r_0=2.57\,\mathrm{\mu m}$ (a.2). \textbf{b)} Phases $\phi_{D,A}$ and amplitudes $|\Omega_{D,A}|$ of the time-optimal gate 13 of a BB code, with $B_{DA}/\Omega_{\rm max}=1.4$ and $B_{AA}/\Omega_{\rm max}=2.24$ (b.1), together with the corresponding data--ancilla (dark line) and ancilla--ancilla (light line) interaction potentials for principal quantum numbers $n_{\rm Rb}=76$ and $n_{\rm Cs}=79$ and lattice spacing $r_0=2.57\,\mathrm{\mu m}$ (b.2).}
    \label{fig:bb_gate}
\end{figure*}

\subsubsection{Gate performance and robustness \label{sec.gate_bb}}

Bivariate Bicycle codes require the implementation of long-range CZ gates between non-nearest-neighbor data and ancilla atoms [configuration (ii) as defined in Sec. \ref{sec.Hamiltonian}]. To mediate interactions between atoms separated by more than $4.5\,r_0$, an additional ancilla atom is introduced. There are eight possible configurations of three-qubit bus gates (see Table ~\ref{tab:bb_gate_frequencies}), each associated with a corresponding time-optimal pulse shape.

As representative examples, we present implementations of the shortest (gate type 6) and longest (gate type 13) gates in panels~\hyperref[fig:bb_gate]{(a.1)}  and~\hyperref[fig:bb_gate]{(b.1)} of Fig.~\hyperref[fig:bb_gate]{\ref{fig:bb_gate}}, respectively. We consider a lattice spacing $r_0 = 2.57\,\mathrm{\mu m}$ and principal quantum numbers $n_{\rm Rb}=76$ and $n_{\rm Cs}=79$. The corresponding pulse durations are $T^{\rm(1b)}\Omega_{\rm max}=12.2$ and $T^{(1b)}\Omega_{\rm max}=12.8$, with infidelities $1-F=5.9\times10^{-3}$ and $1-F=9.2\times10^{-3}$ for gates 6 and 13, respectively, with spontaneous emission and motional errors contribute almost equally. The motion-robustness requirement can be inserted directly into the optimization, and at the cost of slightly longer pulses the infidelity is reduced to $1-F=5.5\times10^{-3}$ and $1-F=5.6\times10^{-3}$, respectively, with now the decay as dominant error source.

The remaining pulse shapes can be generated analogously, and their infidelities are reported in Table~\ref{tab:bb_gate_frequencies}. Gates 1--5, which do not involve mediators, are implemented using a standard blockade scheme with $\Omega_{\rm max}=2\pi\times2\,\mathrm{MHz}$ and sufficiently large $n_{\rm Rb,Cs}$ to ensure $\Omega_{\rm max}\ll B$. Gates 6--13 instead involve one mediator and are realized with $\Omega_{\rm max}=2\pi\times2\,\mathrm{MHz}$, $r_0=2.57\,\mathrm{\mu m}$, $n_{\rm Rb}=76$, and $n_{\rm Cs}=79$. For the most frequent gates, motional robustness constraints are included in the optimization to reduce the weighted average infidelity. The gate infidelity can be further reduced by increasing both $n_{\rm Rb,Cs}$ and $\Omega_{\rm max}$. As an example, we optimized gate 13 using $n_{\rm Rb}=n_{\rm Cs}=92$, $r_0=2.41\,\mathrm{\mu m}$, and $\Omega_{\rm max}=2\pi\times15\,\mathrm{MHz}$, obtaining an infidelity of only $1-F=7.9\times10^{-4}$. \rsub{A complete list of the pulses optimized for long-range stabilizer readout in BB code, together with their parameters, is provided in Table~\ref{tab:all_BB_gates}.} 

We can evaluate the viability of the atomic bus design by computing the average error for the 13 gate types in Table \ref{tab:bb_gate_frequencies}. Weighting each gate by its relative usage frequency we find an average infidelity of $\langle 1-F\rangle = 3.4\times 10^{-3}$. This is lower than the pseudo threshold of $6.5\times 10^{-3}$ reported in \cite{Bravyi2024}. We expect the gate errors could be further reduced with additional optimization and with higher Rydberg levels. We leave a full simulation of this code, accounting for the individual gate errors for future work. 

\subsubsection{Bivariate bicycle layout}

We expect both sensitivity to errors and the time required to execute to scale with the number of bus qubits used. We perform a qubit layout search using the same strategy as \cite{Poole2025a} with some alterations to match the new problem. The initial planar folding scheme from \cite{Poole2025a} results in a layout where stabilizer and data qubits are clumped in a way that does not favorably map onto the bus gate model. We therefore modify the scheme by treating a $2\times 2$ block of atoms containing one atom from each of the four groups as the unit cell for the folding operation. Data qubits are split into two groups $L$ and $R$ with by assigning physical qubits to the left and right half of the parity-check matrices $H_X$ and $H_Z$, respectively. We search over initial positions of the $L$ and $R$ data qubits prior to folding exactly as in \cite{Poole2025a}. The $X$ and $Z$ stabilizer qubits are at this point not yet assigned a location and we must search for configurations that minimize the worst-case number of bus qubits involved in any given $CZ$ operation. In this search, we constrain $X$ and $Z$ qubits to positions that retain the initial unit-cell structure such that any local $2\times 2$ will have one atom of each type. We can build a bipartite graph where the first set of nodes are the unassigned stabilizer qubits and the second are candidate lattice sites. An edge is added to the graph if for a given lattice position, all six stabilizer operations the qubit is involved in have a distance less than some upper bound $D_{\rm max}$. If a perfect matching on this graph is found, the constraint is satisfied and $D_{\rm max}$ is decremented until matching fails. After identifying the smallest satisfiable $D_{\rm max}$, we choose the matching satisfying $D_{\rm max}$ we then determine for each of the stabilizer operations whether a bus qubit is necessary. 

\begin{figure}[ht]
    \centering
    \includegraphics[width=1\linewidth]{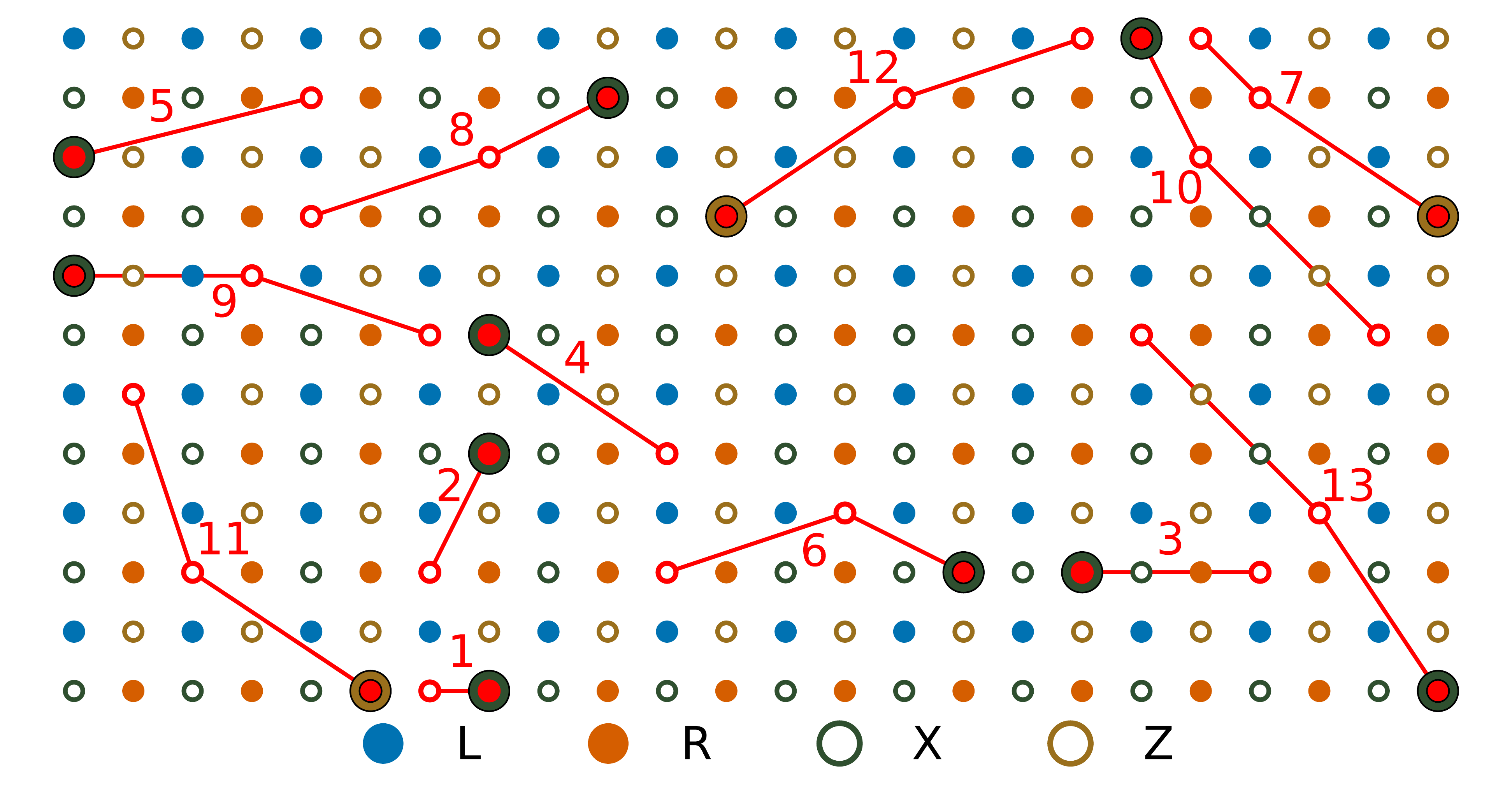}
    \caption{Physical qubit layout for the Bivariate Bicycle code. Qubits connected by red lines serve as a representative example of each gate geometry from Table \ref{tab:bb_gate_frequencies}. All necessary stabilizer operations may be implemented utilizing at most one bus qubit.}
    \label{fig:bb_fig}
\end{figure}

For the $[[144,12,12]]$ code, a layout optimization yields a maximum data--stabilizer separation of $D_{\rm max}=7.81$, measured in units of the lattice spacing. This is larger than the value $D_{\rm max}=7.21$ reported in \cite{Poole2025a}, owing to the additional constraint that every local $2\times2$ qubit plaquette must contain exactly one $X$, $Z$, $L$, and $R$ qubit. This constraint is introduced to reduce the number of distinct gate geometries required by the circuit. To ensure that all stabilizer operations can be implemented using at most one bus qubit, we impose a maximum separation of $4.25$ lattice spacings between adjacent qubits involved in a gate. Indeed, even at this limit, the longest-range operation (gate 13) requires only a single bus qubit, as the data--bus separation is $4.24$ lattice spacings [see Table~\ref{tab:bb_gate_frequencies}]. One final layout  matching is performed which optimizes the average Rydberg interaction strength between pairs of qubits subject to $D_{\rm max}$ to maximize the number of short range gates. This matching can be computed using any linear sum assignment solver. In a gate where it is necessary to assign a bus qubit, we have options for which qubit to assign as the bus. We choose the qubit of the appropriate type which is nearest to the midpoint of the control and target locations. The resulting layout has 13 unique gate geometries. The gate geometries that require bus qubits each have a representative example illustrated in Fig.~\ref{fig:bb_fig}. The frequencies and qubit spacing for each of the gate geometries are listed in Table \ref{tab:bb_gate_frequencies}. 

\subsubsection{Bivariate bicycle gate scheduling \label{sec.bb_scheduling}}

The full syndrome extraction circuit is divided into two sub-circuits. The first will perform all X stabilizer operations using cleanly prepared Z bus qubits. Then, all stabilizer qubits are measured and cleanly prepared X bus qubits for the Z stabilizer operations. Some of the stabilizer operations within each sub-circuit may be performed in parallel. For each operation, an interaction strength is assigned to each qubit in the operation which is equal to $S_{i}=\sum_{j\ne i}1/R_{ij}^6$ where $R_{ij}$ is the distance between qubits $i$ and $j$. For each qubit $i$ involved in the operation, we add other qubits $j$ to the exclusion list of the operation if $\frac{1}{R_{ij}^6 S_{i}} > 0.01.$ Two gates A and B may be performed in parallel if they have the same gate geometries and the exclusion list for each gate does not include the involved qubits of the other gate. Gates are grouped into time slices following a simple heuristic involving the intersection and union of their exclusion lists. We prioritize gates that are compatible but which have the largest intersection of excluded qubits. If multiple gates have the same size intersection, we break ties by considering which option has the smallest union of their exclusion lists. We continue this process for each time slice until no further gates can be added, at which point we start the process over with a new time slice. 

Time slices continue to be scheduled in this way until all operations have been scheduled. We find that all stabilizer operations may be performed in as few as 364 (523 if we change the ratio limit to 0.001) total time slices. We define the compute time as the combined time cost of each slice and readout of the stabilizer qubits. The time cost of each slice is the combined illumination time and switching overhead time. The illumination time cost of each gate type is enumerated in Table \ref{tab:bb_gate_frequencies}, leading to a total illumination time of $723~\mu$s over 364 slices. If the switching time is taken to be $1~\mu$s, the total cost of all time slices is $1087~\mu$s. Without loss of generality, we may assume the $X$ qubits are initially used to form buses for the $Z$ qubits as the two halfs of the circuit are interchangeable. We must perform a readout on the $Z$ qubits in this case prior to using them to form buses for the $X$ qubits in the latter half of the circuit. We must therefore measure the stabilizer qubits twice. Assuming each measurement operation requires $100~\mu$s, this leads to a total cycle time of $1287~\mu$s for the syndrome extraction circuit. This estimate is similar to that presented without bus atoms in \cite{Poole2025a}, but with a more convenient larger lattice spacing
(2.57 vs. $1.7 ~\mu$m) and with lower Rydberg levels (76/79 vs. 50/83/90). The cycle time estimate is also about a factor of two shorter than implementations based on atom motion \cite{Viszlai2025}.

\begin{table}[]
    \resizebox{0.95\columnwidth}{!}{
\begin{tabular}{c|c|c|c|c|c|c}
\toprule
& & \multicolumn{3}{|c|}{Distances} &  &\\ 
Gate Type & Frequency & Stabilizer-Data & Stabilizer-Bus & Data-Bus & $1-F$ & $T(\mathrm{\mu s})$ \\
\hline\hline
1         & 84        & 1.00       &              &             & $1.3\times10^{-3}$ & $0.61$\\
2         & 104       & 2.24       &              &             & $1.3\times10^{-3}$ & $0.61$\\
3         & 34        & 3.00       &              &             & $1.3\times10^{-3}$ & $0.61$\\
4         & 72        & 3.61       &              &             & $1.3\times10^{-3}$ & $0.61$\\
5         & 10        & 4.12       &              &             & $1.3\times10^{-3}$ & $0.61$\\
6         & 38        & 5.00       & 2.24         & 3.16        & $5.5\times10^{-3}$ & $1.01$\\
7         & 12        & 5.00       & 3.61         & 1.41        & $1.1\times10^{-2}$ & $0.81$\\
8         & 62        & 5.39       & 2.24         & 3.16        & $4.8\times10^{-3}$ & $0.98$\\
9         & 164       & 6.08       & 3.00         & 3.16        & $3.5\times10^{-3}$ & $0.88$\\
10        & 9         & 6.40       & 2.24         & 4.24        & $1.1\times10^{-2}$ & $0.96$\\
11        & 9         & 6.40       & 3.61         & 3.16        & $5.5\times10^{-3}$ & $1.00$\\
12        & 154       & 6.71       & 3.61         & 3.16        & $3.8\times10^{-3}$ & $0.96$\\
13        & 112       & 7.81       & 3.61         & 4.24        & $5.6\times10^{-3}$ & $1.07$\\
\hline
Total     & 864       &            &              &             & $3.4\times10^{-3}$ & $723$ \\       
\bottomrule
\end{tabular}}
\caption{Table of frequencies and qubit distances for the 13 different $CZ$ gate types. Each gate type corresponds to a unique geometric configuration of the three atoms, defined by the qubit spacings of pairs of qubits involved in the gate. Distances are expressed in units of the lattice spacing which is set to  $r_0=2.57~\mu\rm m$. Times provided are gate illumination times.}
\label{tab:bb_gate_frequencies}
\end{table}

\section{Code bridging and universality \label{sec.universality}}

In this section, we briefly outline how one can interface high-rate quantum LDPC code memories with surface code units for fault-tolerant quantum computation. The Clifford gate set of the surface code can then be augmented to universality using magic state injection and distillation, which ultimately rely on non-Clifford physical operations, code growth, and transversal entangling operations. This shows that long-range stabilizer readout and transversal gates discussed in the previous section serve as the architectural building blocks for a universal quantum computer, where atom shuttling is confined to background continuous operations, while the algorithmic clock time is only affected by static operations. 

We observe that, despite the higher space overheads, magic state distillation is more compatible with a static dual-species neutral atom architecture than fold-transversal cultivation with surface codes~\cite{Sahay2026}. Cultivation requires iterative code growth, where ancilla atoms of the old code get converted into data qubits in the new code. Within a dual-species framework, this dynamic role reassignment requires continuous lattice reconfiguration and atom shuttling. Moreover, fold-transversal cultivation protocols require non-local, three-qubit entangling gates (e.g., CCZ and CSX), imposing severe control challenges \cite{Sahay2026}.

\subsection{Interfacing qLDPC and surface codes}
In order to process the information stored within a high-rate quantum LDPC memory, a common architectural choice consists of isolating individual logical qubits and teleporting their states into surface codes via lattice surgery \cite{QXu2024}. This requires precise knowledge of the physical data qubits in the support of the logical Pauli strings within the LDPC code lattice to interact them with the surface code boundary. For general quantum LDPC codes, an expression for the logical operators can be found via numerical search or, in special cases, derived semi-analytically. While for La-cross codes logical operators have been found to exhibit structured patterns and translation rules \cite{pecorari2025addressablegatebasedlogicalcomputation}, for general quantum LDPC codes they are typically scattered across the lattice. Consequently, the interface dividing the LDPC and surface code blocks must be filled in with extra ancillary \emph{bridging} atoms, in analogy to the atom-bus approach.  

More quantitatively, the lattice surgery step requires performing fault-tolerant joint logical measurements of the $X_L^\text{qLDPC}\otimes X_L^\text{surf.code}$ ($Z_L^\text{qLDPC}\otimes Z_L^\text{surf.code}$) operator  $\mathcal{O}(d)$ times, with code distance $d$. Finally, the logical qubit of the quantum LDPC code is measured in the $Z$ ($X$) basis to teleport its logical state out of the memory block into the surface code. This operation is deterministic up to a Pauli frame correction to be established at decoding time. Once the state is transferred, the logical computation proceeds by using surface code as quantum processing units.

\subsection{Quasi-static magic state factories}
\label{sec.factory}

In the following, we briefly review the mechanics of standard magic state distillation and outline how one can integrate magic state factories into our architecture in a quasi-static manner.

We consider surface codes of distance $d$ and the standard $15$-to-$1$ magic state distillation protocol \cite{Bravyi2005,Fowler_2012}. The procedure starts with state injection, where a single physical qubit is prepared in a non-Clifford state via a physical $T$ rotation. This state is then expanded (or grown) to a distance-$d$ surface code by sequentially measuring the surrounding lattice stabilizers, which yields a raw, low-fidelity logical magic state $\ket{T}_L$ encoded within a surface code patch. 
The second stage involves concatenating $15$ independently injected surface codes with the $15$-qubit quantum Reed-Muller code \cite{steane1996quantumreedmullercodes}, which supports the logical $T$ gate transversally. To fault-tolerantly concatenate the two codes, one needs to measure the four non-local, weight-$8$ $X$-stabilizers of the Reed-Muller code, namely $X_1X_3X_5X_7X_9X_{11}X_{13}X_{15}$, $X_2X_3X_6X_7X_{10}X_{11}X_{14}X_{15}$, $X_4X_5X_6X_7X_{12}X_{13}X_{14}X_{15}$, and $X_8X_9X_{10}X_{11}X_{12}X_{13}X_{14}X_{15}$. This can be achieved by introducing auxiliary surface code ancilla patches, which are entangled to the $15$ surface code data patches using a sequence of $35$ long-range tCNOT gates. The next step is post-selection based on the ancilla measurement outcomes: If the stabilizer measurements detect a logical error and yield a $-1$ outcome, the protocol is stopped and the states are discarded. Instead, if all outcomes are $+1$, $14$ of the surface code patches are measured in the transversal logical basis. This is equivalent to state teleportation, as it projects the surviving $15$th patch into a single purified logical $\ket{T}_L$ state with enhanced fidelity. The entire process is repeated recursively until the target logical fidelity is reached \cite{Bravyi2005,Fowler_2012}.

Because the stabilizers of the Reed-Muller code span eight distant code blocks, the stabilizer readout step requires non-local connectivity across all the $15$ surface code patches. Within our atom-bus architecture for static long-range entanglement, executing these transversal CNOT gates would require optimized spatial layout and highly long-range interactions mediated by several ancilla atoms. While longer atom-busses are in principle possible, with the currently achievable Rabi rates and lattice spacings they would result in long-range gates with relatively low fidelity, above the surface code error-correction threshold.
Alternatively, one can adopt a hybrid architecture that integrates parallel atom shuttling with bus-mediated long-range gates. This compromise is physically justified because magic state distillation factories operate as offline resource generators by continuously buffering purified states for the main processing units without directly bottlenecking the algorithmic clock time. Consequently, the $15$ logical data patches can be arranged in a static, two-row array, while the four ancilla patches are transported along the central interface in a sequential, conveyor-belt fashion. To minimize latency, motional heating, and decoherence, the required tCNOT operations can be executed statically once the shuttling ancilla patch is found adjacent to the target data patch, in the same way as we have discussed in Sec.~\ref{sec:bellpair}.

\section{Discussion \label{sec.discussion}}

In this paper we have analyzed an architecture for FTQC that is primarily static with long range connectivity provided by an atomic Rydberg bus. We provide detailed simulations of non-local syndrome measurements for fault tolerant quantum memory. The Rydberg bus architecture enables  high rate La-cross codes up to distance $d=7$ without atom motion and bivariate bicycle codes with $d=12$, also without atom motion. Logical states are teleported to surface codes for computation using tCNOT gates that can be implemented up to distance $d=11$ in a static architecture. The same approach can in principle be extended to distances as large as $d=25$, by employing higher Rydberg states and stronger driving fields. We also describe how a combination of bus mediated static interactions and parallel atom transport can be used for asynchronous generation of magic states that are teleported to the surface code for logical universality. Atom transport is thereby relegated to asynchronous generation of magic states and replenishment of lost atoms. 

A  comparison of the computational rate for static vs. transport based architectures is strongly dependent on specific parameters and code choices. We provide quantitative estimates of the QEC cycle time for static implementations of the $d=7$ 2-La-cross codes and $d=12$ bivariate bicycle codes, and compare with implementations based on atom transport. We find that the cycle time of the statically implemented La-cross code is an order of magnitude shorter than a transport based implementation. For the bivariate bicycle code the speedup with the static atomic bus architecture is about a factor of two.  

In addition to the time needed for each QEC cycle 
a full accounting of the wall clock time needed for error corrected logical computation should also include the overhead for atom recooling in deep circuits. Architectures that invoke atom motion for syndrome measurements and for logical connectivity will incur more motional heating than the primarily static architecture analyzed here. 
Hotter atoms lead to reduced gate fidelity and increased logical error rates. Counteracting this in deep circuits requires recooling which operates on much slower timescales than logical gate operations. 
 Shuttling based trapped ion processors allocate a large fraction of the run time to recooling \cite{Pino2021} and a similar situation may apply to neutral atom processors that rely extensively on atom motion.  Thus the speedup factor possible with a static  architecture may in practice be substantially larger than just the ratio of QEC cycle times. 
 This observation has motivated our development of a static, or primarily static,  architecture that minimizes the motional heating associated with atom transport. Of course syndrome measurements on atomic qubits may also lead to heating. In order to realize the full benefit of a static architecture, measurements without additional heating, or only minimal heating,  will be necessary as has been demonstrated in recent work \cite{Scott2025}.

 There are several promising directions in which to extend the analysis presented here.  We have analyzed static implementations of the La-cross and  bivariate bicycle codes. Although these provide much higher code rates than the surface code there are many other codes, including new designs with coding rates $>1/2$ \cite{CZhao2026},  that merit analysis for compatibility with the static architecture presented here.  In Sec.~\ref{sec.factory} we outlined  a hybrid architecture for magic state preparation that combines atom shuttling with bus-mediated long-range gates.  We leave a detailed analysis of that approach, with a full accounting of the resource requirements to future work.

\appendix

\section{Rydberg Atom Interactions \label{app.interactions}}

Rydberg interactions  were calculated for atom pairs in $s$ states, with $m_j=1/2$. Interaction potentials were calculated with the Alkali Rydberg Calculator (ARC) \cite{Sibalic2017} in a $\theta=\phi=0$ geometry and numerically summed \cite{walker2008} to yield an effective blockade. Hamiltonians were created for atom pairs described by $\Delta n=\Delta \ell = 4$, $\Delta E_{\rm max}=25$ GHz. A small magnetic field of 1 G was applied.

Considering the finite confinement of an atom in an optical tweezer, it is useful to consider the variation of the blockade $B$ as a function of atom-atom separation distance $r$. Minimizing the derivative $\partial B/\partial r$ minimizes the shot-to-shot variation in blockade strength, and therefore results in increased gate fidelities.
Due to individual channels of low pair state energy, the blockade curve can differ substantially from a $C_3/r^3+C_6/r^6$ functional form.

\section{Subspace decomposition \label{app.subspaces}}

In this Appendix, we describe how the full Hilbert space can be decomposed into smaller subspaces, thereby reducing the computational cost of the numerical optimization. For simplicity, consider an atomic bus with two data atoms at the ends of the chain (positions $i=0$ and $n_A+1$) and $n_A$ ancilla atoms in between, corresponding to configuration (i) of Sec.~\ref{sec.Hamiltonian}. The following discussion, however, can be straightforwardly extended to configuration (ii). The full Hilbert space introduced in Sec.~\ref{sec.Hamiltonian}, of dimension $\dim\mathcal{H}=3^2\times2^{n_A}$, can be decomposed into smaller invariant subspaces according to the initial state of the data atoms, while the ancilla atoms are always initialized in $\ket{11\ldots1}$. This decomposition is analogous to the one commonly employed in the analysis of two-qubit gates~\cite{Levine2019, Jandura2022}. If the initial state of the data atoms is $\ket{00}$, the data atoms are decoupled and the system acquires a phase due only to the evolution of the ancilla atoms. The dimension of this Hilbert subspace is $\dim \mathcal{H}_{00} = 2^{n_A}$, and the corresponding Hamiltonian reads
\begin{equation}
    H_{00} = \sum_{i=1}^{n_A} H_{A}^{(i)} + H_{\rm int}.
\end{equation}
If the initial state of the data atoms is $\ket{01}$ (or $\ket{10}$), one of the data atoms is decoupled. The dimension of this Hilbert subspace is $\dim \mathcal{H}_{01} = 2^{n_A+1}$, and the corresponding Hamiltonian is
\begin{equation}
    H_{01} = H_D^{(n_A+1)} + \sum_{i=1}^{n_A} H_{A}^{(i)} + H_{\rm int}.
\end{equation}
Note that $H_D^{(i)}$ is now a $2\times2$ matrix. The evolution in $\mathcal{H}_{10}$ is, by symmetry, equivalent to that in $\mathcal{H}_{01}$. Finally, if the initial state of the data atoms is $\ket{11}$, all atoms participate in the dynamics. The dimension of this Hilbert subspace is $\dim \mathcal{H}_{11} = 2^{n_A+2}$, and the Hamiltonian is
\begin{equation}
    H_{11} = H_D^{(0)} + H_D^{(n_A+1)} + \sum_{i=1}^{n_A} H_{A}^{(i)} + H_{\rm int}.
\end{equation}

\section{Analytic estimation of the infidelity \label{app.infidelity}}

In this Appendix, we derive analytical estimates of the infidelity induced by spontaneous emission and thermal motion. Consider a Hamiltonian $H = H^{(0)} + \epsilon H^{(1)}$, where $H^{(0)}$ is the unperturbed Hamiltonian, $H^{(1)}$ is the perturbation, and $\epsilon \ll 1$ is a small parameter. Formally, the evolution operator from $t = 0$ to $t = T$ for such a Hamiltonian reads~\cite{Bergonzoni2026}
\begin{equation}
    \label{eq:evolution_op}
    U(0,T) = U^{(0)}(0,T) - i\epsilon \int_0^T dt U^{(0)}(t,T) H^{(1)} U^{(0)}(0,t),
\end{equation}
where $U^{(0)}$ is the evolution operator associated with the unperturbed Hamiltonian $H^{(0)}$. Given a basis set ${\ket{q}}$ of dimension $Q$, the Bell-state fidelity of a gate is evaluated by taking the overlap between the actual final states $U(0,T)\ket{q}$ and the target states $\ket{\psi_q(T)} = U^{(0)}(0,T)\ket{q}$, such that~\cite{Jandura2022}
\begin{align}
    \label{eq:1-F_1}
    1 - F = & 1 - \frac{1}{Q^2} \left| \sum_{q=1}^Q \bra{q} {U^{(0)}}^\dagger(0,T) U(0,T) \ket{q} \right|^2 \\
    \label{eq:1-F_2}
    = & 1 - \frac{1}{Q^2} \left| Q - i\epsilon \int_0^T dt \sum_{q=1}^Q \bra{\psi_q(t)} H^{(1)} \ket{\psi_q(t)} \right|^2,
\end{align}
where in Eq.~\eqref{eq:1-F} we substitute the expression in Eq.~\eqref{eq:evolution_op}.

In the case of spontaneous emission, the noise Hamiltonian is $H^{(1)}_{\rm ryd} = \sum_i \ket{r}\bra{r}_i$, and the perturbative parameter is $\epsilon = -i\Gamma/2$, where $\Gamma$ is the decay rate at room temperature. Defining
\begin{equation}
    \tau_{\rm ryd} = \frac{1}{Q} \sum_{q=1}^Q \int_0^T dt \bra{\psi_q(t)} H^{(1)}_{\rm ryd} \ket{\psi_q(t)}
\end{equation}
as the average time spent in the Rydberg manifold, Eq.~\eqref{eq:1-F_2} yields, for $\Gamma \tau_{\rm ryd} \ll 1$,
\begin{equation}
    (1 - F)_{\rm ryd} \approx \Gamma \tau_{\rm ryd}.
\end{equation}

In the case of atomic motion, the noise Hamiltonian is $H^{(1)}_{\rm mot} = \sum_{i,j} \ket{rr}\bra{rr}_{ij}$, and the perturbative parameter is $\epsilon = \Delta B$, where $\Delta B$ denotes the variation of the blockade potential induced by positional displacements. Defining
\begin{equation}
    \tau_{\rm 2ryd} = \frac{1}{Q} \sum_{q=1}^Q \int_0^T dt \bra{\psi_q(t)} H^{(1)}_{\rm mot} \ket{\psi_q(t)}
\end{equation}
as the average time during which two nearby atoms are simultaneously in Rydberg states---i.e., the time over which the van der Waals interaction is active---Eq.~\eqref{eq:1-F_2} yields, for $\Delta B\tau_{\rm 2ryd} \ll 1$,
\begin{equation}
(1 - F)_{\rm mot} \approx \Delta B^2 \tau_{\rm 2ryd}^2.
\end{equation}

\rsub{
\section{List of the pulses}
A complete list of all the pulses optimized in this work for transversal CZ gates in the surface code and for long-range stabilizer readout in the La-cross and BB codes, together with their parameters, is provided in Tables~\ref{tab:all_surface_code_gates}, \ref{tab:all_lacross_gates}, and~\ref{tab:all_BB_gates}, respectively.} 
\begin{table*}[ht]
    \centering
    {\color{black}
    \begin{tabular}{c|ccccccccc}
        \toprule
        $d$ & $n_A$ & $\Omega_{\rm max}/(2\pi)$ (MHz) & $n_{\rm Rb}$ & $n_{\rm Cs}$ & $r_0$ ($\mu$m)  & $R_{DD}$ ($\mu$m) & $T$ ($\mu$s) & $1-F$ & Figure \\
        \hline\hline
        5  & 1 & 2 & 71 & 70 & 3    & 15 & 0.672 & $3.2\times10^{-3}$ & \hyperref[fig:gate_surface_code]{\ref{fig:gate_surface_code}(a.1)} \\
        7  & 1 & 2 & 75 & 78 & 2.68 & 18.8 & 0.688 & $2.8\times10^{-3}$ & \hyperref[fig:1ancilla_gate]{\ref{fig:1ancilla_gate}(d)}\\
        7  & 1 & 8 & 89 & 89 & 2.65 & 18.6 & 0.168 & $2.6\times10^{-4}$ & - \\
        9  & 1 & 2 & 73 & 76 & 2.62 & 23.6 & 0.654 & $3.1\times10^{-3}$ & \hyperref[fig:gate_surface_code]{\ref{fig:gate_surface_code}(b.1)} \\
        11 & 2 & 2 & 73 & 76 & 2.62 & 28.8 & 0.820 & $5.3\times10^{-3}$ & \hyperref[fig:gate_surface_code]{\ref{fig:gate_surface_code}(c.1)} \\
        25 & 1 & 2 & 99 & 96 & 1.7  & 42.5 & 0.704 & $3.6\times10^{-3}$ & - \\
        \bottomrule
    \end{tabular}
    }
    \caption{\rsub{Summary table collecting the principal numerical results for all surface-code transversal CZ gates optimized in this work, with $d$ denoting the code distance, $n_A$ the number of ancilla atoms in the atomic bus, $\Omega_{\rm max}$ the Rabi frequency, $n_{\rm Rb,Cs}$ the rubidium and cesium principal quantum numbers, $r_0$ the lattice spacing, $R_{DD}$ the total gate range (data-data distance), $T$ the pulse duration, $1-F$ the simulated infidelity including spontaneous emission and thermal motion, and the corresponding figure of the pulse profile, if included in this manuscript. }}
    \label{tab:all_surface_code_gates}
\end{table*}

\begin{table*}[ht]
    \centering
    {\color{black}
    \begin{tabular}{c|ccccccccc}
        \toprule
        $k$ & opt  & $\Omega_{\rm max}/(2\pi)$ (MHz) & $n_{\rm Rb}$ & $n_{\rm Cs}$ & $r_0$ ($\mu$m) & $R_{DA}^{0,2}$ ($\mu$m) & $T$ ($\mu$s) & $1-F$ & Figure \\
        \hline\hline
        2  & TO     & 2 & 71 & 69 & 3    & 9    & 0.781 & $1.5\times10^{-2}$ & \hyperref[fig:lacross_gate]{\ref{fig:lacross_gate}(a.1)} \\
        2  & robust & 2 & 71 & 69 & 3    & 9    & 0.955 & $7.4\times10^{-3}$ & \hyperref[fig:lacross_gate]{\ref{fig:lacross_gate}(a.2)} \\
        3  & TO     & 2 & 78 & 78 & 3    & 15   & 0.859 & $1.0\times10^{-2}$ & \hyperref[fig:lacross_gate]{\ref{fig:lacross_gate}(b.1)} \\
        3  & robust & 8 & 84 & 81 & 1.75 & 8.75 & 0.239 & $7.1\times10^{-4}$ & \hyperref[fig:lacross_gate]{\ref{fig:lacross_gate}(b.4)} \\
        \bottomrule
    \end{tabular}
    }
    \caption{\rsub{Summary table collecting the principal numerical results for all long-range $k$-La-cross stabilizer readout gates optimized in this work, with $k$ denoting the degree of non-locality, option the optimization technique (either time-optimality or thermal-motion robustness), $\Omega_{\rm max}$ the Rabi frequency, $n_{\rm Rb,Cs}$ the rubidium and cesium principal quantum numbers, $r_0$ the lattice spacing, $R_{DA}^{0,2}$ the total gate range (data-ancilla distance), $T$ the pulse duration, $1-F$ the simulated infidelity including spontaneous emission and thermal motion, and the corresponding figure of the pulse profile, if included in this manuscript. All gates have a single ancilla mediator, $n_A=1$. }}
    \label{tab:all_lacross_gates}
\end{table*}

\begin{table*}[ht]
    \centering
    {\color{black}
    \begin{tabular}{c|cccccccc}
        \toprule
        gate & $\Omega_{\rm max}/(2\pi)$ (MHz) & $n_{\rm Rb}$ & $n_{\rm Cs}$ & $r_0$ ($\mu$m) & $R_{DA}^{0,2}$ ($\mu$m) &  $T$ ($\mu$s) & $1-F$ & Figure \\
        \hline\hline
        6  & 2  & 76 & 79 & 2.57 & 12.9 & 1.011 & $5.5\times10^{-3}$ & \hyperref[fig:bb_gate]{\ref{fig:bb_gate}(a.1)} \\
        7  & 2  & 76 & 79 & 2.57 & 12.9 & 0.812 & $1.1\times10^{-2}$ & - \\
        8  & 2  & 76 & 79 & 2.57 & 13.9 & 0.978 & $4.8\times10^{-3}$ & - \\
        9  & 2  & 76 & 79 & 2.57 & 15.6 & 0.883 & $3.5\times10^{-3}$ & - \\
        10 & 2  & 76 & 79 & 2.57 & 16.4 & 0.963 & $1.1\times10^{-2}$ & - \\
        11 & 2  & 76 & 79 & 2.57 & 16.4 & 1.003 & $5.5\times10^{-3}$ & - \\
        12 & 2  & 76 & 79 & 2.57 & 17.2 & 0.963 & $3.8\times10^{-3}$ & - \\
        13 & 2  & 76 & 79 & 2.57 & 20.1 & 1.074 & $5.6\times10^{-3}$ & \hyperref[fig:bb_gate]{\ref{fig:bb_gate}(b.1)} \\
        13 & 15 & 92 & 92 & 2.41 & 18.8 & 0.151 & $7.9\times10^{-4}$ & - \\
        \bottomrule
    \end{tabular}
    }
    \caption{\rsub{Summary table collecting the principal numerical results for all long-range BB code stabilizer readout gates optimized in this work, with gate denoting the specific gate [see Table~\ref{tab:bb_gate_frequencies}], $\Omega_{\rm max}$ the Rabi frequency, $n_{\rm Rb,Cs}$ the rubidium and cesium principal quantum numbers, $r_0$ the lattice spacing, $R_{DA}^{0,2}$ the total gate range (data-ancilla distance), $T$ the pulse duration, $1-F$ the simulated infidelity including spontaneous emission and thermal motion, and the corresponding figure of the pulse profile, if included in this manuscript. All gates have a single ancilla mediator, $n_A=1$. }}
    \label{tab:all_BB_gates}
\end{table*}

\section{Infidelity due to atomic shuttling \label{app.heating}}

In this Appendix, we estimate the impact of atomic shuttling on the gate fidelity through heating and dephasing. The origin of this dephasing is the differential trapping potential $\delta\omega = \omega_0 - \omega_1$ experienced by the qubit states $\{\ket{0}, \ket{1}\}$. Consider a single atom initially prepared at time $t=0$ in the state $\ket{\psi(0)} = \ket{+}_{\rm int} \ket{0}_{\rm vib}$, where the internal state is the superposition $\ket{+}_{\rm int} = (\ket{0}_{\rm int} + \ket{1}_{\rm int})/\sqrt{2}$ and in the vibrational ground state $\ket{0}_{\rm vib}$. When the optical tweezer is accelerated to transport the atom, the vibrational wavefunction acquires a displacement whose magnitude depends on the smoothness of the acceleration, and it evolves into a coherent state
\begin{equation}
    \ket{\alpha}_{\rm vib} = e^{-|\alpha|^2/2} \sum_{n=0}^\infty \frac{\alpha^n}{\sqrt{n!}} \ket{n}_{\rm vib}.
\end{equation}
Here $|\alpha|^2 \equiv \delta n$ is the average number of vibrational excitations induced in the harmonic trap due to the acceleration.

After the acceleration, the shuttling proceeds inertially for a time $t = t^*$. Each state $\ket{j}_{\rm int} \ket{n}_{\rm vib}$ acquires a phase $\exp[-i \omega_j (n + 1/2)t^*]$, where $j = 0,1$ labels the internal qubit state and $n = 0,1,2,\ldots$ labels the vibrational eigenstates. Since the acquired phase depends on both the internal and vibrational degrees of freedom, decoherence is induced. In particular, after a time $t^*$ the state becomes
\begin{align}
    \nonumber
    \ket{\psi(t^*)}  = & \frac{e^{-|\alpha|^2/2}}{\sqrt{2}}  \sum_{n=0}^\infty \frac{\alpha^n}{\sqrt{n!}} \Big(  \ket{0}_{\rm int} \ket{n}_{\rm vib}e^{-i\omega_0\left(n+\frac{1}{2}\right)t^*}\\
    &+ \ket{1}_{\rm int} \ket{n}_{\rm vib}e^{-i\omega_1\left(n+\frac{1}{2}\right)t^*} \Big)
\end{align}

Consider the density matrix $\rho(t^*) = \ket{\psi(t^*)}\bra{\psi(t^*)}$. By tracing out the vibrational degrees of freedom, we obtain the reduced density matrix for the internal space,
\[
\rho_{\rm int}(t^*) = \sum_{m=0}^{\infty} {}_{\rm vib}\!\bra{m}\,\rho(t^*)\,\ket{m}_{\rm vib}.
\]
In this way, one obtains
\begin{align}
    \rho_{\rm int}(t^*) = & \frac{e^{-|\alpha|^2}}{2} \sum_{m=0}^\infty \frac{\alpha^{2m}}{m!} \Big( \ket{0}\bra{0} +  \ket{1}\bra{1}\\
    \nonumber
    & + \ket{0}\bra{1} e^{-i\delta\omega(m+\frac{1}{2})t^*} + \ket{1}\bra{0} e^{i\delta\omega(m+\frac{1}{2})t^*} \Big)
\end{align}
Consider the coherence, i.e., the off-diagonal element of $\rho_{\rm int}(t^*)$. Evaluating the exponential series, we obtain
\begin{equation}
    \rho^{\rm int}_{01}(t^*) = \frac{e^{-\delta n}}{2}\, e^{-i\delta\omega t^*/2}\, e^{\delta n e^{-i\delta\omega t^*}}.
\end{equation}
Assuming $\delta\omega t^* \ll 1$, we can perform a Taylor expansion of the exponential $\exp(-i\delta\omega t^*)\approx1-i\delta\omega t^* - (\delta\omega t^*)^2/2$ and obtain
\begin{equation}
    \rho^{\rm int}_{01}(t^*) \approx \frac{e^{-i\delta\omega\left(\delta n + \frac{1}{2}\right)t^*}}{2}\, e^{-(t^*/T_2)^2},
\end{equation}
where the complex exponential arises from coherent evolution, while the second exponential represents a Gaussian decay of the coherence, with decay time
\begin{equation}
T_2 = \sqrt{\frac{2}{\delta n\, \delta\omega^2}}.
\end{equation}

The fidelity of a quantum memory experiment can be evaluated as
\begin{equation}
    F(t^*) = {}_{\rm int}\bra{+} \rho_{\rm int}(t^*) \ket{+}_{\rm int}
    \approx 1 - \frac{1}{2}\left(\frac{t^*}{T_2}\right)^2.
\end{equation}

As an estimate for the shuttling time $t^*$, we consider the duration of a minimal-jerk trajectory~\cite{Chinnarasu2025, Carruthers1965},
\begin{equation}
    t_{\rm mj} = \frac{\sqrt{2}\,\sqrt[3]{15R}}{\sqrt[6]{\delta n}\,\sqrt[3]{x_{\rm ho}}\,\omega} \propto d^{1/3},
\end{equation}
where $R = d\,r_0$ is the total displacement and $x_{\rm ho}=1/\sqrt{2m\omega}$ is the harmonic oscillator length of the trap, with $m$ the atomic mass. For realistic parameters, this yields an infidelity scaling
\begin{equation}
    (1 - F)_{\rm shuttle}(d) \approx 6\times10^{-4}\, d^{2/3},
\end{equation}
for the choice $\delta n=0.1$, $\omega=2\pi\times100\,\mathrm{kHz}$, $\delta\omega/\omega\approx6.9\times10^{-3}$, $r_0=3\,\mathrm{\mu m}$ and $x_{\rm ho}=35\,\mathrm{nm}$, \rsub{and with a time $t_{\rm mj}=d^{1/3}\times36\,\mathrm{\mu s}$.}

\rsub{We note that the above discussion applies to hyperfine qubits, consistently with our choice of Rb and Cs states. However, for nuclear-spin qubits, both qubit states should experience the same trapping potentials, i.e. $\delta\omega=0$, and the dephasing and bit-flip error rates associated with atomic transport are likely to be negligible. Nevertheless, minimizing qubit errors during transport is also critical in a reconfigurable processor with nuclear-spin qubits, due to the associated leakage and loss. In particular, in Ref.~\cite{Zhang2026}, where nuclear-spin qubits are encoded in the metastable ${}^3{\rm P}_0$ state of $^{171}$Yb, transport errors dominate the current circuit implementation, accounting for $66\%$ of the logical error rate, while also introducing a substantial temporal overhead.\\
Regarding the motional time $t^*$, nuclear-spin encoding also relaxes the constraint imposed by qubit coherence on transport speed. However, faster atom movements can introduce additional heating and motional excitations, potentially resulting in increased atom loss. Increasing the AOD trap depth can help suppress motional heating, but at the same time increases photon-scattering rates. Moreover, a critical aspect of atomic transport is the ramping of the AOD/SLM trap depth required to hand off qubits from one trap to another, which, for example, accounts for $1.6\,\mathrm{ms}$ in \cite{Zhang2026}.}

\begin{figure*}
    \centering
    \includegraphics[width=\linewidth]{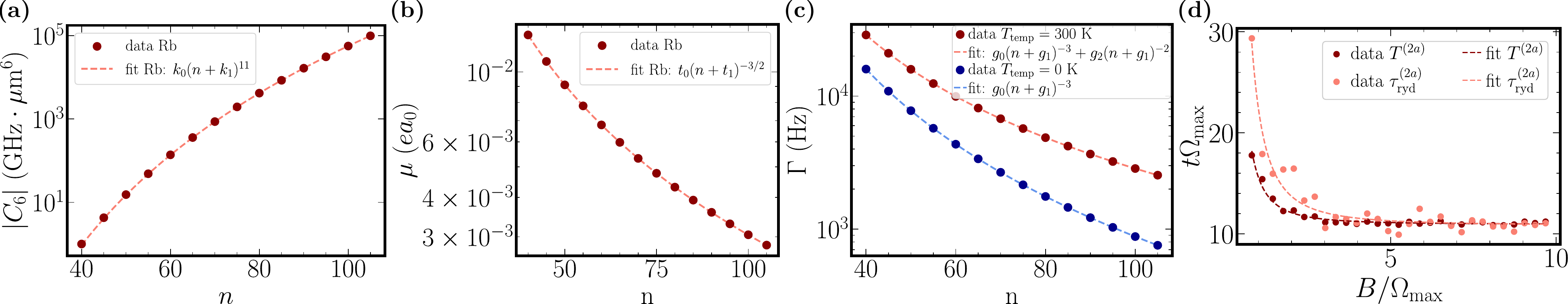}
    \caption{\textbf{(a)} Dispersion coefficient $C_6$ as a function of the principal quantum number $n$ in rubidium. Fit parameters: $k_0\approx-1.0\times10^{-17}\,\mathrm{GHz\cdot\mu m^6}$ and $k_1\approx-5.1$. \textbf{(b)} Dipole matrix element $t$ as a function of $n$ in rubidium. Fit parameters: $t_0\approx2.9\,ea_0$ and $t_1\approx-3.6$. \textbf{(c)} Decay rate $\Gamma$ as a function of $n$ in rubidium at cryogenic temperature $T_{\rm temp}=0\,\mathrm{K}$ and room temperature $T_{\rm temp}=300\,\mathrm{K}$. Fit parameters: $g_0\approx8.0\times10^8\,\mathrm{Hz}$ and $g_1\approx-3.1$, and $g_0\approx7.6\times10^8\,\mathrm{Hz}$, $g_1\approx-2.9$ and $g_2\approx1.9\times10^7\,\mathrm{Hz}$, respectively. \textbf{(d)} Gate duration $T^{\rm(2a)}$ and time spent in the Rydberg manifold $\tau_{\rm ryd}^{\rm(2a)}$ for time-optimal pulses with two mediators, shown as functions of $B=B_{AA}=B_{DA}$. Numerical optimization results (dots) are fitted using Eq.~\eqref{eq:tau_ryd_generic} (dashed lines).}
    \label{fig:appendix}
\end{figure*}

\section{Scalability of the gates \label{app:scalability}}

In this Appendix, we derive an analytical estimate for the fidelity scaling of symmetric bus gates in terms of the main physical parameters, namely the number of ancilla mediators $n_A$, the principal quantum number $n$, the laser electric field $E$, the lattice spacing $r_0$, and the surface-code distance $d$. Consider a bus gate with $n_A$ mediating ancilla atoms implementing a transversal CZ gate between two surface codes of distance $d$. Assuming equally spaced atoms, the distance between neighboring atoms involved in the gate is $R=r_0d/(n_A+1)$, where $r_0$ is the lattice spacing. The vdW interaction between nearby atoms is $B=C_6/R^6$. For simplicity, in this qualitative estimate we do not distinguish between inter- and intra-species interactions. The dispersion coefficient scales as $C_6=k_0(n+k_1)^{11}$, where $n$ is the principal quantum number of the Rydberg state and $k_0$, $k_1$ are numerical constants. For rubidium, for example, $k_0\approx-1.0\times10^{-17}\,\mathrm{GHz\cdot\mu m^6}$ and $k_1\approx-5.1$ [Fig.~\hyperref[fig:appendix]{\ref{fig:appendix}(a)}].

The Rabi frequency driving the gate is $\Omega=E\mu/\hbar$, where $E$ is the laser electric-field amplitude and $\mu$ is the dipole matrix element associated with the transition $\ket{1}\to\ket{n}$,
\begin{equation}
    \mu=e\bra{1}\vec{r}\cdot\hat{\epsilon}\ket{n},
\end{equation}
with $e$ the electron charge, $\vec{r}$ the electron position operator, and $\hat{\epsilon}$ the laser polarization. The dipole matrix element scales approximately as $\mu=\mu_0(n+\mu_1)^{-3/2}$, with suitable constants $\mu_0$ and $\mu_1$. For rubidium, one finds $\mu_0\approx2.9\,ea_0$ and $\mu_1\approx-3.6$ [Fig.~\hyperref[fig:appendix]{\ref{fig:appendix}(b)}].

The dominant contribution to the gate infidelity arises from spontaneous emission, characterized by a decay rate scaling as $\Gamma=g_0(n+g_1)^{-3}$, where $g_0$ and $g_1$ are numerical constants. For rubidium at $T_{\rm temp}=0\,\mathrm{K}$, $g_0\approx8.0\times10^8\,\mathrm{Hz}$ and $g_1\approx-3.1$ [Fig.~\hyperref[fig:appendix]{\ref{fig:appendix}(c)}]. At room temperature, $T_{\rm temp}=300\,\mathrm{K}$, accounting also for black-body-radiation induced spontaneous emission the scaling is better represented by $\Gamma=g_0(n+g_1)^{-3}+g_2(n+g_1)^{-2}$ \cite{Beterov2009}, with similar parameters $g_0\approx7.6\times10^8\,\mathrm{Hz}$, $g_1\approx-2.9$ and $g_2\approx1.9\times10^8\,\mathrm{Hz}$.

Combining the previous results, the ratio between the interaction strength and the Rabi frequency can be estimated as
\begin{equation}
    \frac{B}{\Omega} = \frac{\hbar k_0 (n_A+1)^6 (n+k_1)^{11}(n+\mu_1)^{3/2}}{E \mu_0 (r_0 d)^6}.
\end{equation}

In the main text, the dependence of $\tau_{\rm ryd}^{\rm(1a)}$ on $B/\Omega$ was determined numerically for the case of a single ancilla mediator ($n_A=1$), see Eq.~\eqref{eq:tau_ryd_1a}. Extending this analysis to longer buses with multiple mediating ancilla atoms is more involved, since additional interaction scales enter the problem, notably the data--ancilla interaction $B_{DA}$ and the ancilla--ancilla interaction $B_{AA}$. Moreover, the numerical complexity of the optimal-control optimization grows rapidly with the number of atoms.

Nevertheless, we estimated the scaling of the time spent in the Rydberg manifold for a two-ancilla bus, $\tau_{\rm ryd}^{\rm(2a)}$, assuming $B_{DA}=B_{AA}$ and neglecting next-nearest-neighbor interactions. We found the same qualitative functional dependence observed in the single-ancilla case [see Fig.~\hyperref[fig:1ancilla_gate]{\ref{fig:1ancilla_gate}(b.1)}]. Motivated by this result, we infer that for a symmetric bus with $n_A$ ancilla mediators,
\begin{equation}
    \label{eq:tau_ryd_generic}
    \tau_{\rm ryd} \approx \frac{1}{\Omega} \left[ \tau_0(n_A) + \frac{\tau_1(n_A)}{(B/\Omega)^2} \right],
\end{equation}
where $\tau_0(n_A)$ and $\tau_1(n_A)$ are numerical coefficients depending on the number of ancilla atoms. Explicitly, for $n_A=1$ we find $\tau_0(1)=5.2$ and $\tau_1(1)=3.4$ [Eq.~\eqref{eq:tau_ryd_1a} and Fig.~\hyperref[fig:1ancilla_gate]{\ref{fig:1ancilla_gate}(b.1)}], while for $n_A=2$ we obtain $\tau_0(2)=10.8$ and $\tau_1(2)=11.6$ [Fig.~\hyperref[fig:appendix]{\ref{fig:appendix}(d)}].

We can finally estimate the gate infidelity $1-F$, dominated by spontaneous emission, as
\begin{align}
    \label{eq:1-F_scaling_full}
    & 1-F \approx \frac{\hbar \left[g_0 (n+g_1)^{-3} + g_2(n+g_1)^{-2}\right] (n+\mu_1)^{3/2}}{E \mu_0} \\
    \nonumber
    & \times\left( \tau_0(n_A) + \tau_1(n_A)
    \frac{E^2 \mu_0^2 (r_0d)^{12}}{\hbar^2 k_0^2 (n_A+1)^{12} (n+k_1)^{22} (n+\mu_1)^3} \right).
\end{align}
This expression allows us to estimate the scalability of the bus gates as a function of the physical parameters. In general, reducing the lattice spacing $r_0$ and increasing the principal quantum number $n$, although both introduce additional experimental challenges, are beneficial for reaching larger code distances $d$ at fixed infidelity. Stronger laser fields $E$ can improve the fidelity, but they introduce a tradeoff between faster gate operation (the term scaling as $1/E$) and increased sensitivity in the weak-blockade regime (the term scaling as $E$). However, stronger driving is generally beneficial when combined with larger principal quantum numbers $n$. A simplified version of Eq.~\eqref{eq:1-F_scaling_full} is presented in the main text in Eq.~\eqref{eq:1-F_scaling}, assuming $n_A=1$, $n+k_1\approx n+\mu_1\approx n+g_1$ and $g_2\ll g_0$.

\section{Simulation of motion}

In this Appendix, we detail how thermal motional errors are included in our numerical simulations. Thermal atomic motion affects the interatomic distance $R$ and therefore the vdW interaction strength $B$, reducing the fidelity of bus gates that operate at finite $B$. Before excitation to the Rydberg manifold, the atoms are confined in optical tweezers of trapping frequency $\omega$. Assuming a thermal distribution of vibrational states in the harmonic trap, the position uncertainty $\sigma_\alpha=\sqrt{\langle x_\alpha^2\rangle-\langle x_\alpha\rangle^2}$ for an atom of species $\alpha=D,A$ is
\begin{equation}
    \sigma_\alpha = \sqrt{\frac{\hbar}{2m_\alpha\omega}\coth\!\left(\frac{\hbar\omega}{2k_B T_{\rm th}}\right)},
\end{equation}
where $m_\alpha$ is the atomic mass, $k_B$ the Boltzmann constant, and $T_{\rm th}$ the temperature.

The relative fluctuation of the interatomic distance between two atoms of species $\alpha_1$ and $\alpha_2$ is estimated as
\begin{equation}
    \Delta R = \frac{\sqrt{\sigma_{\alpha_1}^2+\sigma_{\alpha_2}^2} }{R}.
\end{equation}
Using the distance dependence of the vdW interaction, $B(R)$, we then estimate the corresponding relative fluctuation of the interaction strength as
\begin{equation}
    \Delta B =\max\left\{\frac{|B(R)-B(R(1\pm\Delta R))|}{B(R)}
    \right\}.
\end{equation}
Finally, the resulting gate infidelity is estimated by replacing $B$ with $B(1+\Delta B)$ in the Hamiltonian and reevaluating the gate fidelity.

In the numerical simulations, we assume a trapping frequency $\omega=2\pi\times100\,\mathrm{kHz}$ and a temperature $T_{\rm th}=5\times10^{-6}\,\mathrm{K}$, corresponding to interatomic distance fluctuations of approximately $40$--$50\,\mathrm{nm}$.

\begin{acknowledgments}
This research has received funding from the European Union’s Horizon Europe Research and Innovation Programme under the HORIZON-CL4-2021-DIGITAL-EMERGING-01-30 via the project 101070144 (EuRyQa),
and from the French National Research Agency under the Investments of the Future Program projects ANR-21-ESRE-0032 (aQCess), 
ANR-17-EURE-0024 (QMat), 
and ANR-22-CMAS-0001 France 2030 (QuanTEdu-France).
The work in Madison was supported by ARO under contract W911NF2410382 and the US National Science Foundation under Award 2016136 for the QLCI center Hybrid Quantum Architectures and Networks.
Computing time was provided by the High-Performance Computing Center of the University of Strasbourg and the Center for High Throughput Computing (CHTC) at University of Wisconsin--Madison. Part of the computing resources were funded by the Equipex Equip@Meso project (Programme Investissements d'Avenir) and the CPER Alsacalcul/Big Data.
\end{acknowledgments}

\bibliography{qc_refs,optics,saffman_refs,rydberg,atomic,thispaper}

\end{document}